\documentstyle[12pt,a4]{article}
\input epsf
\global\arraycolsep=2pt

\begin{document}

\begin{titlepage}

\begin{flushright}
CERN-TH.7524/94\\
hep-ph/9502264\\
revised April 1995
\end{flushright}

\vspace{0.5cm}

\begin{center}
\Large\bf Resummation of Renormalon Chains for\\
Cross Sections and Inclusive Decay Rates
\end{center}

\vspace{1.0cm}

\begin{center}
Matthias Neubert\\
{\sl Theory Division, CERN, CH-1211 Geneva 23, Switzerland}
\end{center}

\vspace{1.2cm}

\begin{abstract}
Recently, we have developed a formalism to evaluate QCD loop diagrams
with a single virtual gluon using a running coupling constant at the
vertices. This corresponds to an all-order resummation of certain
terms (the so-called renormalon chains) in a perturbative series and
provides a generalization of the scale-setting prescription of
Brodsky, Lepage and Mackenzie. In its original form, the method is
applicable to Green functions without external gluons and to
euclidean correlation functions. Here we generalize the approach to
the case of cross sections and inclusive decay rates, which receive
both virtual and real gluon corrections. We encounter
nonperturbative ambiguities in the resummation of the perturbative
series, which may hinder the construction of the operator product
expansion in the physical region. The origin of these ambiguities and
their relation to renormalon singularities in the Borel plane is
investigated by introducing an explicit infrared cutoff. The ratios
$R_{e^+ e^-}$ and $R_\tau$ are discussed in detail.
\end{abstract}

\vspace{1.0cm}

\centerline{(Submitted to Physical Review D)}

\vspace{2.0cm}

\noindent
CERN-TH.7524/94\\
April 1995

\end{titlepage}

\section{Introduction}

Recently, an approach has been developed to analyse the momentum flow
in Feynman diagrams containing a single virtual gluon line
\cite{part1}. It was initiated by the observation that sometimes the
``typical'' momenta in a loop diagram are different from the
``natural'' scale of the process. This is indicated by the fact that
in the perturbative series for some quantities depending on a single
mass scale $M$, there remain large two-loop corrections of order
$\beta_0\,\alpha_s^2$ when one uses the ``natural'' scale $M$ to
evaluate the running coupling constant. Here $\beta_0 =
11-\frac{2}{3}\,n_f$ is the first coefficient of the
$\beta$-function, and $n_f$ denotes the number of light quark
flavours. Brodsky, Lepage and Mackenzie (BLM) have argued that the
appearance of such terms indicates an inappropriate choice of the
renormalization scale, and that one should eliminate them by
readjusting the scale in the one-loop running coupling constant
\cite{BLM,LeMa}. This prescription defines the BLM scale $\mu_{\rm
BLM}$, which may be interpreted as the ``typical'' scale of virtual
momenta in Feynman diagrams. One of the goals of Ref.~\cite{part1}
was to understand the relation between the BLM scale and the
``natural'' scale of a process, in particular in cases where
$\mu_{\rm BLM}\ll M$.

Consider the calculation of a physical (i.e.\ renormalization-scheme
invariant and infrared finite) quantity $S(M^2)$ at order $\alpha_s$
in perturbation theory. The BLM prescription is equivalent to using
the average virtuality of the gluon as the scale in the running
coupling constant. In Ref.~\cite{part1} we have generalized this
proposal by performing the calculation with a running coupling
constant $\alpha_s(-k^2)$ at the vertices, where $k$ is the momentum
flowing through the virtual gluon line. This idea is not new (see,
for instance, the discussion in Ref.~\cite{LeMa}), but because of its
computational complexity it had not been pursued previously to the
point of practical implementation. The result of such a calculation,
which gives the average of the running coupling constant over the
virtual momenta in Feynman diagrams, may be written as (for
simplicity we normalize the lowest-order contribution to unity)
\begin{equation}\label{resum}
   S_{\rm res}(M^2) = 1 + \int\limits_0^\infty\!{\rm d}\tau\,
   \widehat w(\tau)\,{\alpha_s(\tau e^C M^2)\over 4\pi} \,,
\end{equation}
where the scheme-independent function $\widehat w(\tau)$ describes
the distribution of virtualities in the loop calculation. The
constant $C$ depends on the renormalization scheme in such a way that
the value of the coupling constant $\alpha_s(\tau e^C M^2)$ is
scheme-independent (in MS-like schemes). This implies that the
product
\begin{equation}\label{LamVdef}
   \Lambda_V \equiv e^{-C/2}\,\Lambda_{\rm QCD}
\end{equation}
is scheme-independent, where $\Lambda_{\rm QCD}$ is the scale
parameter in the one-loop expression for the running coupling
constant. We note that $C=-5/3$ in the $\overline{\rm MS}$ scheme,
$C=-5/3+\gamma-\ln 4\pi$ in the MS scheme, and $C=0$ in the so-called
V scheme, in which the running coupling constant is defined in terms
of the heavy-quark potential \cite{LeMa}. Although the V scheme is
particularly convenient for our considerations, we shall display the
dependence on $C$ in order to keep the discussion general.

Eq.~(\ref{resum}) is equivalent to an all-order resummation of
certain terms in the perturbative series for the quantity $S(M^2)$.
This explains the subscript ``res''. Using the one-loop expression
for the running coupling constant to relate $\alpha_s(\tau e^C M^2)$
to $\alpha_s(M^2)$, one finds that
\begin{equation}\label{whatint}
   S_{\rm res}(M^2) = 1 + \sum_{n=1}^\infty \Bigg(
   {\alpha_s(M^2)\over 4\pi} \Bigg)^n\,c_n\,\beta_0^{n-1} \,,
\end{equation}
where the coefficients $c_n$ are given by the integrals
\begin{equation}
   c_n = \int\limits_0^\infty\!{\rm d}\tau\,\widehat w(\tau)\,
   (-C-\ln\tau)^{n-1} \,.
\end{equation}
In other words, our approach resums all terms of order
$\beta_0^{n-1}\alpha_s^n$ in the perturbative series for the quantity
$S(M^2)$. The resummation of such terms has also been considered by
Beneke and Braun \cite{BBnew} (see also \cite{xxx}), using however a
different formalism. To understand what it corresponds to, consider a
Feynman diagram with a single virtual gluon line. A particular set of
higher-order graphs is obtained by inserting $(n-1)$ light-quark
loops on the gluon propagator. The resulting contributions are of
order $n_f^{n-1}\alpha_s^n$. In an abelian theory they are obviously
related to the renormalization of the coupling constant. In a
non-abelian theory one may try to incorporate the effects of gauge-
and ghost-field loops by replacing $n_f$ by $-\frac{3}{2}\,\beta_0$.
After this replacement, these contributions are no longer related in
an obvious way to subclasses of diagrams; however, they are called
renormalon chains. It is the all-order resummation of renormalon
chains that is accomplished in (\ref{whatint}).

It is instructive to relate the approximation $S_{\rm res}(M^2)$ to
the BLM scale-setting prescription \cite{BLM}. We find
\begin{eqnarray}\label{BLMrel}
   S_{\rm res}(M^2) &=& 1 + N\,{\alpha_s(\mu_{\rm BLM}^2)\over\pi}\,
    \Bigg\{ 1 + \Delta\,\Bigg( {\beta_0\,\alpha_s(\mu_{\rm BLM}^2)
    \over 4\pi} \Bigg)^2 + \dots \Bigg\} \nonumber\\
   &=& S_{\rm BLM}(M^2) + {N\Delta\over 16\pi^3}\,\beta_0^2\,
    \alpha_s^3(\mu_{\rm BLM}^2) + \dots \,,
\end{eqnarray}
where
\begin{eqnarray}\label{Deltadef}
   N &=& {1\over 4}\,\int\limits_0^\infty\!{\rm d}\tau\,
    \widehat w(\tau) \,, \nonumber\\
   \mu_{\rm BLM}^2 &=& \exp\Big( \langle \ln\tau \rangle + C
    \Big)\,M^2 \,, \nonumber\\
   \phantom{ \Bigg[ }
   \Delta &=& \sigma^2 = \langle \ln^2\!\tau \rangle
    - \langle \ln\tau \rangle^2 \,.
\end{eqnarray}
We use the symbol
\begin{equation}
   \langle f(\tau) \rangle =
   {\int\limits_0^\infty\!{\rm d}\tau\,\widehat w(\tau)\,
    f(\tau) \over \int\limits_0^\infty\!{\rm d}\tau\,
    \widehat w(\tau)}
\end{equation}
for the average of a function $f(\tau)$ over the distribution
$\widehat w(\tau)$. The one-loop coefficient $N$, the value of the
coupling constant $\alpha_s(\mu_{\rm BLM}^2)$, and the parameter
$\Delta$ are renormalization-scheme invariant. The first correction
to the BLM scheme appears at order $\beta_0^2\,\alpha_s^3$ and is
related to the width (with respect to $\ln\tau$) of the distribution
function.

Let us now investigate the infrared (IR) properties of the integral
representation (\ref{resum}). The fact that the integration extends
to $\tau=0$ indicates the appearance of nonperturbative effects,
which arise because of the divergent behaviour of perturbative
expansions in QCD. Because of a factorial growth of the expansion
coefficients $c_n$, a perturbative series such as the one on the
right-hand side in (\ref{whatint}) has a vanishing radius of
convergence. It is useful to introduce a generating function for
these coefficients,
\begin{equation}\label{Sudefi}
   \widehat S(u) = e^{C u} \sum_{n=1}^\infty
   {u^{n-1}\over\Gamma(n)}\,c_n \,,
\end{equation}
which can be identified with the Borel transform of the series
(\ref{whatint}) with respect to the inverse coupling constant
\cite{tHof}, in the limit where $\beta_0\to\infty$. The series on the
right-hand side of (\ref{Sudefi}) is usually convergent in a finite
interval $u\in\,]\!-\!j,k\,[$ around the origin, where $j$ and $k$
are called the positions of the nearest UV and IR renormalons (see
below). Note that we have defined the function $\widehat S(u)$ in a
scheme-independent way. Its relation to the distribution function
$\widehat w(\tau)$ is~\cite{part1}
\begin{eqnarray}\label{Swrela}
   \widehat S(u) &=& \int\limits_0^\infty\!{\rm d}\tau\,
    \widehat w(\tau)\,\tau^{-u} \,, \nonumber\\
   \widehat w(\tau) &=& {1\over 2\pi i}
    \int\limits_{u_0-i\infty}^{u_0+i\infty}\!{\rm d}u\,
    \widehat S(u)\,\tau^{u-1} \,.
\end{eqnarray}
The choice of the parameter $u_0$ in the inverse Mellin
representation is arbitrary as long as it is inside the interval
$]\!-\!j,k\,[$, where the first integral is well defined. Outside
this interval the Borel transform $\widehat S(u)$ is defined by
analytic continuation. The fact that the function $\widehat w(\tau)$
is well-defined and has a concrete physical interpretation, whereas
the definition of $\widehat S(u)$ requires an analytic continuation,
will become important later.

The Borel transformation can be inverted to give
\begin{eqnarray}\label{Laplace}
   S_{\rm Borel}(M^2) &=& 1 + {1\over\beta_0}\,
    \int\limits_0^\infty\!{\rm d}u\,\widehat S(u)\,e^{-C u}\,
    \exp\bigg( -{4\pi u\over\beta_0\,\alpha_s(M^2)} \bigg)
    \nonumber\\
   &=& 1 + {1\over\beta_0}\,\int\limits_0^\infty\!
    {\rm d}u\,\widehat S(u)\,\bigg( {\Lambda_V^2\over M^2}
    \bigg)^u \,,
\end{eqnarray}
where $\Lambda_V$ is the scheme-independent parameter defined in
(\ref{LamVdef}). If the integral existed, it would define the Borel
sum of the partial series on the right-hand side of (\ref{whatint}).
In general, however, the Borel transform $\widehat S(u)$ contains
singularities on the real $u$-axis, and the result of the integration
depends on how these singularities are regulated. Much of the
nonperturbative structure of QCD can be inferred from a study of the
Borel transform \cite{tHof}--\cite{Chris}. Its singularities on the
negative axis arise from the large-momentum region in Feynman
diagrams and are called ultraviolet (UV) renormalons. They are Borel
summable and pose no problem to performing the integral in
(\ref{Laplace}). The singularities on the positive axis arise from
the low-momentum region in Feynman diagrams and are called IR
renormalons. Their presence leads to an ambiguity in the evaluation
of the Borel integral, reflecting the fact that in (\ref{whatint})
one is attempting to sum up a series which is not Borel summable.

In our approach, IR renormalon ambiguities appear because the
$\tau$-integral in (\ref{resum}) runs over the Landau pole in the
running coupling constant. In order to regularize the integral, one
may write
\begin{equation}\label{regul}
   {\beta_0\,\alpha_s(\tau e^C M^2)\over 4\pi}
   = {1\over\ln M^2/\Lambda_V^2 + \ln\tau}
   = {\rm P}\,\bigg( {1\over\ln\tau-\ln\tau_L} \bigg)
   + \eta\,\delta(\ln\tau-\ln\tau_L) \,,
\end{equation}
where $\tau_L=\Lambda_V^2/M^2$ is the position of the Landau pole,
``P'' denotes the principle value, and $\eta$ is a complex parameter,
which depends on the regularization prescription. A particularly
useful regularization is to set $\eta=i\pi$, corresponding to
\begin{equation}\label{ieps}
   {\beta_0\,\alpha_s(\tau e^C M^2)\over 4\pi}
   = {1\over\ln\tau-\ln\tau_L-i\epsilon} \,.
\end{equation}
Let us show that the resummation (\ref{resum}) together with this
regularization prescription is equivalent to performing the Borel
integral (\ref{Laplace}) along a contour above the real $u$-axis.
We note that
\begin{eqnarray}\label{proof}
   S_{\rm res}(M^2) &=& 1 + {1\over\beta_0}\,
    \int\limits_0^\infty\!{\rm d}\tau\,\widehat w(\tau)\,
    {1\over\ln\tau - \ln\tau_L - i\epsilon} \nonumber\\
   &=& 1 + {i\over\beta_0}\,\int\limits_0^\infty\!{\rm d}\tau\,
    \widehat w(\tau)\,\int\limits_0^\infty\!{\rm d}\sigma\,
    \bigg( {\tau\over\tau_L} \bigg)^{-i\sigma}\,
    e^{-\epsilon\sigma} \nonumber\\
   &=& 1 + {1\over\beta_0}\,\int\limits_0^{i\infty}\!{\rm d}u\,
    \widehat S(u)\,\bigg( {\Lambda_V^2\over M^2} \bigg)^u\,
    e^{i\epsilon u} \,,
\end{eqnarray}
where we have used the integral representation (\ref{Swrela}) of the
Borel transform $\widehat S(u)$ in terms of the distribution function
$\widehat w(\tau)$. Note that interchanging the order of integrations
when passing from the second to the third line is justified, since
both integrals converge. Let us now assume that the Borel transform
$\widehat S(u)$ is an analytic function in the complex $u$-plane
apart from singularities on the real axis. It is then allowed to
rotate the integration contour in the last integral in (\ref{proof})
into a contour that runs above the real axis, and then to take the
limit $\epsilon\to 0$. This proves the equivalence of our resummation
with the Borel integral in (\ref{Laplace}). Using the same argument
it follows that choosing a $+i\epsilon$ to regulate the Landau pole
in (\ref{ieps}) corresponds to performing the Borel integral below
the real $u$-axis, and regulating the Landau pole with a principle
value prescription is equivalent to taking the principle value of the
Borel integral.

The dependence on the regularization prescription leads to an
intrinsic ambiguity in the definition of $S_{\rm res}(M^2)$.
Inserting (\ref{regul}) into (\ref{resum}) and defining the
renormalon ambiguity $\Delta S_{\rm ren}$ as the coefficient of
$\eta$, we find
\begin{equation}\label{DeltaS}
   \Delta S_{\rm ren} = {\tau_L\over\beta_0}\,\widehat w(\tau_L)
   \simeq {w_0\over\beta_0}\,\bigg( {\Lambda_V^2\over M^2}
   \bigg)^k \,,
\end{equation}
where we have used the fact that $\tau_L\ll 1$ to expand the
distribution function:
\begin{equation}\label{wtauexp}
   \widehat w(\tau) = w_0\,\tau^{k-1} + \dots \quad
   \mbox{for $\tau\to 0$.}
\end{equation}
It is the asymptotic behaviour of the function $\widehat w(\tau)$ for
small values of $\tau$ that determines the size of the renormalon
ambiguity. The power $k$ coincides with the position of the nearest
IR renormalon pole in the Borel plane \cite{part1}. Note that $k>0$
in order for the integral in (\ref{resum}) to be IR convergent.

In Ref.~\cite{part1} we have developed the resummation technique for
QCD Green functions without external gluon fields and for euclidean
correlation functions of currents. In these cases only virtual gluon
corrections have to be considered. Both from the phenomenological and
from the conceptual point of view it is interesting to extend the
formalism to the case of cross sections and inclusive decay rates,
which receive both virtual and real gluon corrections. This
generalization is the subject of the present work. In particular, we
shall consider the perturbative series for the cross section
$\sigma(e^+ e^-\to\mbox{hadrons})$ and for the decay rate
$\Gamma(\tau\to\nu_\tau+ \mbox{hadrons})$. In the $\overline{\rm MS}$
scheme, one finds at order $\alpha_s^2$ \cite{Ree1}--\cite{Rtau2}
\begin{eqnarray}\label{examples}
   {1\over 3\,(\sum Q_i^2)}\,{\sigma(e^+ e^-\to\mbox{hadrons})
   \over\sigma(e^+ e^-\to\mu^+\mu^-)}
   &=& 1 + {\alpha_s(s)\over\pi}
    + (0.17\beta_0 + 0.08)\,\Bigg( {\alpha_s(s)\over\pi}
    \Bigg)^2 \,, \nonumber\\
   {1\over 3}\,{\Gamma(\tau\to\nu_\tau+\mbox{hadrons})\over
   \Gamma(\tau\to\nu_\tau\,e\,\bar\nu_e)}
   &=& 1 + {\alpha_s(m_\tau^2)\over\pi}
    + (0.57\beta_0 + 0.08)\,\Bigg( {\alpha_s(m_\tau^2)\over\pi}
    \Bigg)^2 \,,
\end{eqnarray}
where $Q_i$ denote the electric charges of the quarks, $s$ is the
centre-of-mass energy, and quark mass effects are neglected. The
terms proportional to $\beta_0$ give the main contribution to the
two-loop coefficients. In the case of $R_\tau$ they are numerically
quite important. If one uses the BLM prescription to absorb these
terms into a redefinition of the scale in the running coupling
constant, one obtains $\mu_{\rm BLM}^{e^+ e^-}\simeq 0.71\sqrt{s}$
and $\mu_{\rm BLM}^\tau\simeq 0.32\,m_\tau$ (in the $\overline{\rm
MS}$ scheme), respectively. A more striking example is provided by
the parton model prediction for the semileptonic decay rate
$\Gamma(b\to u\,e\,\bar\nu_e)$. The corresponding perturbative series
at two-loop order is \cite{LSW}
\begin{equation}
   {\Gamma(b\to u\,e\,\bar\nu_e)\over\Gamma_{\rm tree~level}}
   = 1 - 2.41\,{\alpha_s(m_b^2)\over\pi}
   - (3.22\beta_0 + k)\,\Bigg( {\alpha_s(m_b^2)\over\pi}
   \Bigg)^2 \,,
\end{equation}
where the constant $k$ is yet unknown. The corresponding BLM scale is
$\mu_{\rm BLM}^{b\to u}\simeq 0.07\,m_b$. Here and in the case of
$R_\tau$, the BLM scales are so low that one may doubt the
reliability of the perturbative expansion. It is therefore important
to understand the origin of these low scales and to consider
higher-order terms in the series. The fact that the contributions of
order $\beta_0\,\alpha_s^2$ in (\ref{examples}) give the dominant
two-loop corrections justifies, to some extent, that we concentrate
on the class of higher-order contributions provided by renormalon
chains. Clearly, the resummation in (\ref{whatint}) does not replace
an exact higher-order calculation, but nevertheless it provides a
nontrivial all-order resummation of a gauge-invariant subset of
corrections. As such, it seems worth while to apply this approach in
the cases considered above. However, quantities defined in the
physical region (i.e.\ the region of time-like momenta) differ in two
respects from those considered in Ref.~\cite{part1}, which were
defined in the euclidean region. First, in the calculation of
radiative corrections both virtual and real gluons contribute, and
only the sum of their contributions is IR finite \cite{Kino,LeeN}.
Clearly, in such a situation one has to generalize the idea of
performing a one-loop calculation with a running coupling constant
$\alpha_s(-k^2)$, since $k^2=0$ for real gluons. Second, the operator
product expansion (OPE), which provides the framework for a
systematic analysis of nonperturbative effects, can be justified in
the euclidean region only, although it is sometimes argued that a
``generalized OPE'' based on quark--hadron duality holds in the
physical region after applying a ``smearing procedure'' \cite{smear}.
Below we shall discuss in detail the resummation of renormalon chain
contributions for the ratios $R_{e^+ e^-}$ and $R_\tau$, encountering
some new features arising from the above-mentioned complications. The
``linear'' form of the integral representation in (\ref{resum}) is
replaced by ``non-linear'' representations, in which instead of the
coupling constant there appears a process-dependent function of the
coupling constant. The choice of this function is not unique, leading
to nonperturbative ambiguities in the definition of the resummed
series that are not related to IR renormalons. Based on our results
we cannot exclude that the one-to-one correspondence between the size
of long-distance contributions and the position of renormalon
singularities, which in the euclidean region establishes the link
between the OPE and the Borel integral, may be lost in the physical
region.

In Sect.~\ref{sec:2} we review some results derived in
Ref.~\cite{part1} for the perturbative series for the correlator of
two vector currents in the euclidean region. We discuss the
connection between IR renormalons and nonperturbative effects. By
introducing a factorization scale, which serves to separate short-
and long-distance contributions, we construct the OPE to order
$1/(Q^2)^2$. In Sect.~\ref{sec:3} we perform the analytic
continuation to the physical region and discuss the resummation of
renormalon chains for the cross-section ratio $R_{e^+ e^-}$.
Section~\ref{sec:4} is devoted to the analogous discussion for the
decay-rate ratio $R_\tau$. In Sect.~\ref{sec:add} we investigate the
origin of the nonperturbative ambiguities in the definition of the
resummed series. We study long-distance effects and their relation to
the singularities in the Borel plane by introducing an IR cutoff to
regularize the Borel integral. A numerical analysis of the results is
presented in Sect.~\ref{sec:5}. In Sect.~\ref{sec:6} we summarize our
results and point out some phenomenological implications. The
relation of our approach to the resummation procedure proposed by
Beneke and Braun \cite{BBnew} is discussed in the Appendix.

\section{Current correlator in the euclidean region}
\label{sec:2}

\begin{figure}[htb]
   \vspace{0.5cm}
   \epsfysize=6cm
   \centerline{\epsffile{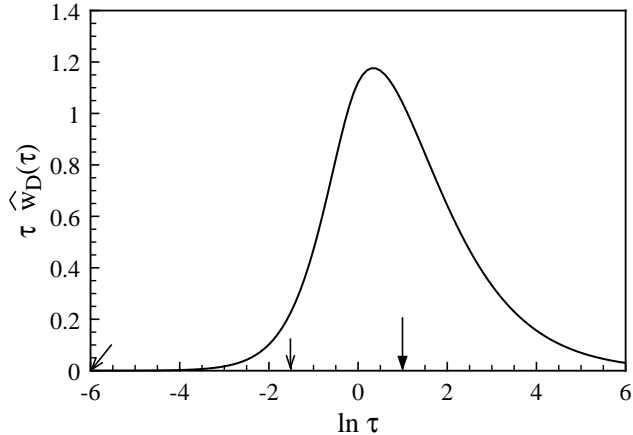}}
   \centerline{\parbox{13cm}{\caption{\label{fig:1}
The distribution function $\tau\,\widehat w_D(\tau)$ as a function of
$\ln\tau$. The long arrow indicates the average value of $\ln\tau$,
which determines the BLM scale. The short arrows show the point
$\tau=\lambda^2/Q^2$ for $\lambda=1$~GeV and $Q^2=m_\tau^2$ (right)
and $(20~\mbox{GeV})^2$ (left).}}}
\end{figure}

Consider the correlator $\Pi(Q^2)$ of two vector currents $j^\mu=\bar
q\,\gamma^\mu q$ in the euclidean region ($Q^2=-q^2>0$):
\begin{equation}
   i \int\!{\rm d}^4 x\,e^{i q\cdot x}\,\langle\,0\,|\,
   T\{ j^\mu(x),j^\nu(0) \}\,|\,0\,\rangle
   = (q^\mu q^\nu - q^2 g^{\mu\nu})\,\Pi(Q^2) \,.
\end{equation}
For simplicity we shall neglect the quark masses. The momentum
transfer $Q^2\gg\Lambda_V^2$ provides the large mass scale. The
derivative of $\Pi(Q^2)$ with respect to $Q^2$ is UV
convergent. We define
\begin{equation}
   D(Q^2) = 4\pi^2 Q^2\,{{\rm d}\Pi(Q^2)\over{\rm d}Q^2} \,.
\end{equation}
In Ref.~\cite{part1} we have constructed the resummation of
renormalon chains for the corresponding perturbative series. The
result is
\begin{equation}\label{dQ2}
   D_{\rm res}(Q^2) = 1 + \int\limits_0^\infty\!{\rm d}\tau\,
   \widehat w_D(\tau)\,{\alpha_s(\tau e^C Q^2)\over 4\pi} \,,
\end{equation}
where the distribution function is given by
\begin{eqnarray}\label{wDfun}
   \widehat w_D(\tau) &=& 8 C_F\,\bigg\{
    \bigg( {7\over 4} - \ln\tau \bigg)\,\tau
    + (1+\tau)\,\Big[ L_2(-\tau) + \ln\tau\,\ln(1+\tau) \Big]
    \bigg\} \,;\quad \tau<1 \,, \nonumber\\
   && \nonumber\\
   \widehat w_D(\tau) &=& 8 C_F\,\bigg\{ 1 + \ln\tau
    + \bigg( {3\over 4} + {1\over 2}\,\ln\tau \bigg)\,{1\over\tau}
    \nonumber\\
   &&\qquad \mbox{}+ (1+\tau)\,\Big[ L_2(-\tau^{-1}) - \ln\tau\,
    \ln(1+\tau^{-1}) \Big] \bigg\} \,;\quad \tau>1 \,.
\end{eqnarray}
Here $C_F=4/3$ is a colour factor, and $L_2(x)=-\int_0^x {{\rm
d}y\over y}\ln(1-y)$ is the dilogarithm function. The function
$\widehat w_D(\tau)$ and its first three derivatives are continuous
at $\tau=1$, but higher derivatives are not. A graphical
representation of the distribution function is shown in
Fig.~\ref{fig:1}. We find it most useful to show the product
$\tau\,\widehat w_D(\tau)$ as a function of $\ln\tau$, since then the
integrals $\langle\ln^n\!\tau\rangle$ have a direct graphical
interpretation. In order to associate mass scales with the
$\tau$-values in the figure, we note that $\ln(\mu^2/Q^2)=\ln\tau+C$,
where $\mu$ is the scale in the running coupling constant. In the V
scheme, the point $\ln\tau=0$ corresponds to $\mu^2=Q^2$; in the
$\overline{\rm MS}$ scheme, it corresponds to $\ln\tau=5/3$. We
observe that the distribution function is rather narrow and centred
around $\ln\tau\simeq 1$. The long arrow indicates the average value
$\langle\ln\tau\rangle$. In order to calculate the BLM scale and the
width parameter $\Delta$ defined in (\ref{Deltadef}), we compute
\begin{eqnarray}
   N &=& {3\over 4}\,C_F = 1 \,, \nonumber\\
    \langle \ln\tau \rangle &=& 4\zeta(3) - {23\over 6}
    \simeq 0.975 \,, \nonumber\\
   \phantom{ \bigg[ }
   \langle \ln^2\!\tau \rangle &=& 18 - 12\zeta(3)\simeq 3.575 \,,
\end{eqnarray}
where $\zeta(3)\simeq 1.20206$. This leads to
\begin{eqnarray}
   {\mu_{\rm BLM}^D\over\sqrt{Q^2}} &\simeq& 1.628\,e^{C/2}
    ~\stackrel{\overline{\rm MS}}{\to}~ 0.708 \,, \nonumber\\
   \Delta_D &\simeq& 2.625 \,,\qquad \sigma_D \simeq 1.620 \,.
\end{eqnarray}
Note that both in the $\overline{\rm MS}$ and in the V scheme, the
BLM scale is close to the ``natural'' scale $\sqrt{Q^2}$.

The asymptotic behaviour of the distribution function for $\tau\to 0$
is
\begin{equation}\label{asy}
   \widehat w_D(\tau) = C_F\,\Big\{ 6\tau
   + (4\ln\tau - 6)\,\tau^2 + O(\tau^3) \Big\} \,.
\end{equation}
As mentioned in the introduction, the linear term signals that the
nearest IR renormalon singularity in the Borel transform of the
perturbative series is located at $u=2$. Indeed, in the case of the
function $D(Q^2)$ the Borel transform is \cite{Broa,Bene}
\begin{equation}\label{SDu}
   \widehat S_D(u) = {32 C_F\over 2-u}\,\sum_{k=2}^\infty\,
   {(-1)^k\,k\over\big[ k^2-(1-u)^2\big]^2}
   = {6 C_F\over 2-u} + \dots \,,
\end{equation}
where the ellipses represent terms that are regular at $u=2$. The
presence of IR renormalons leads to an ambiguity in the value of the
resummed series, which according to (\ref{DeltaS}) is given by
\begin{equation}\label{Dren}
   \Delta D_{\rm ren} = {\Lambda_V^2\over\beta_0\,Q^2}\,
   \widehat w_D(\Lambda_V^2/Q^2) \simeq {6 C_F\over\beta_0}\,
   \bigg( {\Lambda_V^2\over Q^2} \bigg)^2 \,.
\end{equation}

The appearance of IR renormalons acts as a reminder that the result
of any perturbative calculation in QCD is incomplete; it must be
supplemented by nonperturbative corrections. Only the sum of all
perturbative and nonperturbative contributions is unambiguous.
Unlike any finite-order calculation, the representation (\ref{resum})
makes explicit that perturbative calculations contain long-distance
contributions from the region of low momenta in Feynman diagrams.
Moreover, it provides a convenient way to implement Wilson's
construction of the OPE \cite{Wils}. Since the integration variable
$\tau$ can be interpreted as a physical scale parameter, one can
separate the contributions from different momentum scales by
splitting up the integral into a short- and a long-distance piece. We
define
\begin{eqnarray}\label{separ}
   D_{\rm res}(Q^2) &=& 1 + \int\limits_{\lambda^2/Q^2}^\infty\!
    {\rm d}\tau\,\widehat w_D(\tau)\,
    {\alpha_s(\tau e^C Q^2)\over 4\pi}
    + \int\limits_0^{\lambda^2/Q^2}\!{\rm d}\tau\,
    \widehat w_D(\tau)\,{\alpha_s(\tau e^C Q^2)\over 4\pi}
    \nonumber\\
   \phantom{ \bigg[ }
   &\equiv& 1 + D_{\rm sd}(Q^2,\lambda^2)
    + D_{\rm ld}(Q^2,\lambda^2) \,.
\end{eqnarray}
Here $\lambda$ acts as a factorization (or separation) scale, which
should be chosen such that $\Lambda_V<\lambda\ll\sqrt{Q^2}$. The
short-distance contribution can be reliably calculated in
perturbation theory, and it is free of renormalon ambiguities. The
long-distance contribution must be combined with other
nonperturbative corrections. Only the sum of all long-distance
contributions is well defined. The dependence on the arbitrary scale
$\lambda$ must cancel in the final result. This $\lambda$-dependence
can be controlled in perturbation theory by means of the
renormalization-group equation
\begin{equation}\label{lamRGE}
   \lambda^2\,{{\rm d}\over{\rm d}\lambda^2}\,
   D_{\rm ld}(Q^2,\lambda^2)
   = -\lambda^2\,{{\rm d}\over{\rm d}\lambda^2}\,
   D_{\rm sd}(Q^2,\lambda^2) \nonumber\\
   = {\alpha_s(e^C \lambda^2)\over 4\pi}\,
   {\lambda^2\over Q^2}\,\widehat w_D(\lambda^2/Q^2) \,.
\end{equation}
If $\lambda^2/Q^2\ll 1$, the $Q^2$-dependence of the long-distance
contribution is determined by the asymptotic behaviour of the
distribution function for small values of $\tau$ and coincides with
the $Q^2$-dependence of the renormalon ambiguity. In the present
case, it follows that $D_{\rm ld}(Q^2,\lambda^2)\propto 1/(Q^2)^2$.
At the same order in the OPE there appear nonperturbative effects
parametrized by the gluon condensate \cite{SVZ}, which has an UV
renormalon ambiguity that compensates the ambiguity from the IR
renormalon in the perturbative series \cite{reno2}--\cite{Bene}.
Hence, to this order we may write for the physical correlator
\begin{eqnarray}
   D(Q^2) &=& 1 + D_{\rm sd}(Q^2,\lambda^2)
    + D_{\rm ld}(Q^2,\lambda^2) + {2\pi\over 3(Q^2)^2}\,
    \langle\alpha_s\,G^2\rangle + \dots \nonumber\\
   &\equiv& 1 + D_{\rm sd}(Q^2,\lambda^2) + {2\pi\over 3(Q^2)^2}\,
    \langle\alpha_s\,G^2\rangle(\lambda) + \dots \,,
\end{eqnarray}
where the last equation defines a scale-dependent gluon condensate,
which is free of renormalon ambiguities. From (\ref{lamRGE}), it
follows that
\begin{equation}
   \lambda^2\,{{\rm d}\over{\rm d}\lambda^2}\,
   \langle\alpha_s\,G^2\rangle(\lambda)
   = {9 C_F\over 4\pi^2}\,\alpha_s(e^C\lambda^2)\,\lambda^4 \,.
\end{equation}

Let us now present a numerical analysis of our results. We choose
$Q^2=m_\tau^2$ as the momentum scale. In Table~\ref{tab:1} we compare
the following approximations for $D(m_\tau^2)$: the one- and
partial\footnote{To be consistent, we take into account only the part
of the two-loop corrections proportional to $\beta_0\,\alpha_s^2$.
For the function $D(m_\tau^2)$ the remaining two-loop correction is
$0.08\,(\alpha_s/\pi)^2\simeq 8\times 10^{-4}$.}
two-loop expressions evaluated using the ``natural'' scale $Q^2$ in
the running coupling constant, the one-loop expression evaluated
using the BLM scale, the truncated series including the first
correction to the BLM scheme (denoted by $D_{\rm BLM^*}$) given by
the term proportional to $\Delta$ in (\ref{BLMrel}), and the partial
resummation of the series provided by the integral over the
distribution function regulated with the principle value
prescription. We use the one-loop expression for the running coupling
constant with $n_f=3$ light quark flavours. We take
$\alpha_s(m_\tau^2)=0.32$ in the $\overline{\rm MS}$ scheme,
corresponding to $\Lambda_3^{\overline{\rm MS}}\simeq 201$~MeV. The
corresponding value of the scale parameter in the V scheme is
$\Lambda_V\simeq 461$~MeV. We observe that the effect of higher-order
corrections is not negligible. The resummation of renormalon chains
leads to a 3.3\% increase in the value of $D(m_\tau^2)$ with respect
to the two-loop result. This effect is twice as big as the two-loop
correction itself. Most of the higher-order corrections are taken
into account if one includes the first correction to the BLM scheme.
A nice way to compare the all-order resummation of renormalon chains
with the BLM approximation is to define a scale $\mu_*$ such that the
one-loop correction evaluated at this scale reproduces the resummed
series, i.e.\ $D_{\rm res}(m_\tau^2)\equiv 1 +
\alpha_s(\mu_*^2)/\pi$. We find $\mu_*/\mu_{\rm BLM}\simeq 0.69$. The
fact that $\mu_*<\mu_{\rm BLM}$ means that higher-order corrections
effectively decrease the average virtuality.

\begin{table}[t]
\centerline{\parbox{15cm}{\caption{\label{tab:1}
Comparison of various approximations for the euclidean correlator
$D(Q^2)$ at $Q^2=m_\tau^2$.}}}
\vspace{0.5cm}
\centerline{\begin{tabular}{cccccc}
\hline\hline
\rule[-0.2cm]{0cm}{0.7cm} $D_{\rm 1~loop}$ & $D_{\rm 2~loop}$ &
 $D_{\rm BLM}$ & $D_{\rm BLM^*}$ & $D_{\rm res}$ &
 $\Delta D_{\rm ren}$ \\
\hline
\rule{0cm}{0.6cm} 1.102 & 1.118 & 1.121 & 1.145 & 1.151 &
 $3.3\times 10^{-3}$ \\
\hline\hline
\end{tabular}}
\vspace{0.5cm}
\end{table}

\begin{figure}[htb]
   \vspace{0.5cm}
   \epsfysize=6cm
   \centerline{\epsffile{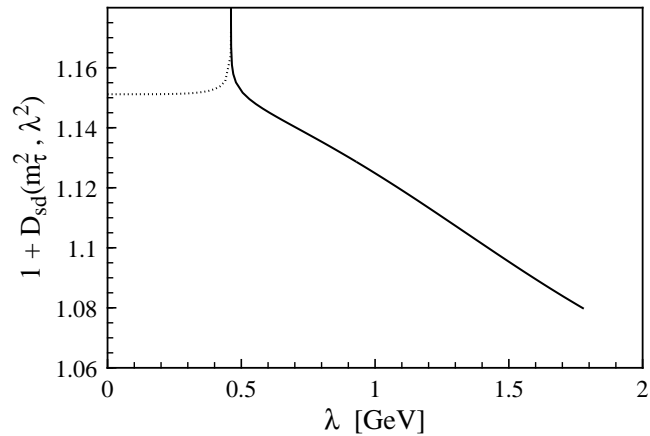}}
   \centerline{\parbox{13cm}{\caption{\label{fig:2}
Short-distance contribution to $D(m_\tau^2)$ as a function of the
factorization scale. The dotted line is used in the region below the
Landau pole, where the short-distance contribution is no longer well
defined.}}}
\end{figure}

As mentioned above, the problem of IR renormalons can be avoided by
introducing a factorization scale $\lambda$ to separate short- and
long-distance contributions. For the two values
$Q^2=(20~\mbox{GeV})^2$ and $Q^2=m_\tau^2$, the factorization point
$\tau=\lambda^2/Q^2$ for $\lambda=1$~GeV is indicated by the short
arrows in Fig.~\ref{fig:1}.\footnote{In the $\overline{\rm MS}$
scheme, $\lambda=1$~GeV corresponds to the rather low scale
$\mu=e^{-5/6}\lambda\simeq 0.43$~GeV. The value of the ratio
$\lambda/\Lambda_V=\mu/\Lambda_{\rm QCD}$ is scheme independent.}
Whereas long-distance contributions are completely negligible for
$Q^2=(20~\mbox{GeV})^2$, they are important for $Q^2=m_\tau^2$. This
is reflected in Fig.~\ref{fig:2}, where we show the short-distance
contribution $1+D_{\rm sd}(m_\tau^2,\lambda^2)$, defined in
(\ref{separ}), as a function of the factorization scale. The solid
line is used in the range $\Lambda_V<\lambda<m_\tau$, where $D_{\rm
sd}(m_\tau^2,\lambda^2)$ is well defined. The dotted line is used for
the region below the Landau pole. Roughly speaking, perturbation
theory can be trusted down to a value $\lambda_0\simeq 0.93$ GeV
(corresponding to $\mu\simeq 0.40$~GeV in the $\overline{\rm MS}$
scheme), where $\alpha_s(e^C\lambda_0^2)=1$. The variation of the
short-distance contribution with $\lambda$ provides an estimate of
the importance of long-distance effects. In the present case, these
effects are of order a few per cent.

\section{Analytic continuation and $R_{e^+ e^-}$}
\label{sec:3}

\begin{figure}[htb]
   \vspace{0.5cm}
   \epsfysize=6cm
   \centerline{\epsffile{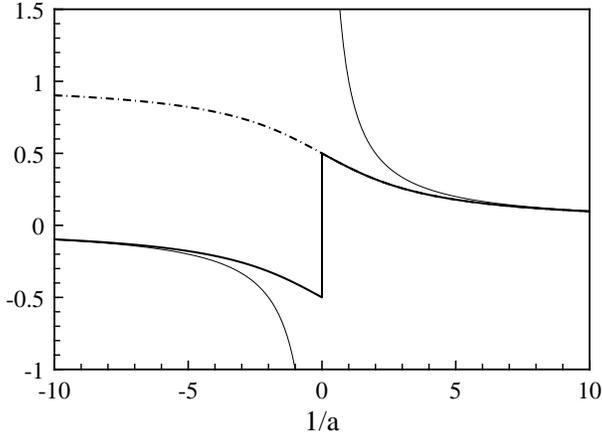}}
   \centerline{\parbox{13cm}{\caption{\label{fig:3}
Effective coupling constants $F_1(a)$ (dash-dotted line) and $F_2(a)$
(solid line) as a function of $1/a$. Both functions coincide for
$a>0$. The thin line shows the bare coupling constant $a$.}}}
\end{figure}

Let us now consider the analytic continuation of the euclidean
correlator $\Pi(Q^2)$ to the physical region, where $s=-Q^2>0$. The
imaginary part of the correlator is related to the total cross
section for the process $e^+ e^-\to\mbox{hadrons}$ at the
centre-of-mass energy $s$ by
\begin{equation}\label{SRdef}
   R_{e^+ e^-}(s) = {\sigma(e^+ e^-\to\mbox{hadrons})
    \over\sigma(e^+ e^-\to\mu^+\mu^-)}
   = 12\pi \Big( \sum Q_i^2 \Big)\,\mbox{Im}\,\Pi(-s+i\epsilon)
   \equiv 3 \Big( \sum Q_i^2 \Big)\,R(s) \,,
\end{equation}
where $Q_i$ denote the electric charges of the quarks. Our goal is to
resum the renormalon chain contributions (i.e.\ terms of order
$\beta_0^{n-1}\alpha_s^n$) in the perturbative series for the
quantity $R(s)$, and to derive a representation of the result as an
integral over a distribution function. To this end, we integrate
(\ref{dQ2}) with respect to $\ln Q^2$ and use the one-loop expression
for the $\beta$-function to obtain \cite{part1}
\begin{equation}
   4\pi^2\,\Big[ \Pi(Q^2) - \Pi(Q_0^2) \Big]_{\rm res}
   = \ln{Q^2\over Q_0^2} + {1\over\beta_0}\,
   \int\limits_0^\infty\!{\rm d}\tau\,\widehat w_D(\tau)\,
   \ln{\alpha_s(\tau e^C Q_0^2)\over\alpha_s(\tau e^C Q^2)} \,.
\end{equation}
Here $Q_0^2$ is an arbitrary subtraction point. Next we perform the
analytic continuation $Q^2\to-s+i\epsilon$, using that for the
one-loop running coupling constant
\begin{equation}
   {1\over\pi}\,\mbox{Im}\,\ln{1\over\alpha_s(-\tau e^C s+i\epsilon)}
   = F_1\big(a(\tau s)\big) \,,
\end{equation}
where
\begin{equation}\label{Fdef}
   F_1(a) = {1\over\pi}\,\mbox{arccot}\bigg( {1\over\pi a} \bigg)
   = {1\over\pi}\arctan(\pi a) + \Theta(-a) \,,
\end{equation}
and
\begin{equation}\label{adef}
   a(\mu^2) \equiv {\beta_0\,\alpha_s(e^C\mu^2)\over 4\pi}
   = {1\over\ln\mu^2/\Lambda_V^2} \,.
\end{equation}
Combining the above results we obtain an integral representation for
the resummed perturbative series $R_{\rm res}(s)$, in which there
appears the same distribution function as in (\ref{dQ2}), but a
non-linear function of the coupling constant. Using the notation
\begin{equation}\label{WeewD}
   W_{e^+ e^-}(\tau) = \widehat w_D(\tau)
\end{equation}
for the distribution function in this ``non-linear'' representation,
we write the result as
\begin{equation}\label{Srep}
   R_{\rm res}^{(1)}(s) = 1 + {1\over\beta_0}\,
   \int\limits_0^\infty\!{\rm d}\tau\,W_{e^+ e^-}(\tau)\,
   F_1\big(a(\tau s)\big) \,.
\end{equation}
The need for the label ``1'' will become clear below. A similar
integral representation has been suggested by Beneke and Braun
\cite{BBnew}. The precise relation between their approach and ours is
clarified in the Appendix. These authors suggest to interpret
$F_1(a)$ as an effective coupling constant, which has the attractive
features that it agrees with the usual coupling constant in the
perturbative region $0<a\ll 1$, but it has a smooth behaviour in the
IR region. In Fig.~\ref{fig:3} we show the effective coupling
constant $F_1(a)$ and the bare coupling constant $a$ as a function of
$1/a=\ln\mu^2/\Lambda_V^2$.

Since the perturbative series for $R(s)$ and $D(Q^2)$ agree to order
$\alpha_s^2$, the BLM scales are the same in the two cases.
Differences appear first at order $\alpha_s^3$, i.e.\ in the value of
the parameter $\Delta$ in (\ref{BLMrel}). In the case of $R(s)$, this
parameter is no longer given by the square of the width $\sigma_D$ of
the distribution function. Taking into account that $F_1(a)=a-\pi^2
a^3/3+O(a^5)$, we find that
\begin{eqnarray}\label{muee}
   {\mu_{\rm BLM}^{e^+ e^-}\over\sqrt{s}} &\simeq& 1.628\,e^{C/2}
    ~\stackrel{\overline{\rm MS}}{\to}~ 0.708 \,, \nonumber\\
   \Delta_{e^+ e^-} &=& \sigma_D^2 - {\pi^2\over 3} \simeq -0.665 \,.
\end{eqnarray}
Since $\Delta_{e^+ e^-}$ is negative, the leading correction to the
BLM scheme effectively increases the value of the scale. In general,
non-linear representations like (\ref{Srep}), for which the growth of
the running coupling constant in the IR region is damped, lead
to an improved convergence of the perturbative series.

Let us now investigate the renormalon ambiguity of the resummed
series (\ref{Srep}). Expanding the function $F_1(a)$ in powers of the
coupling constant and using the relation
\begin{equation}
   \Big[ \pi\,a(\tau s) \Big]^{n+1}
   = {\pi\over n!}\,(-\delta_\tau)^n\,a(\tau s) \,;\quad
   \delta_\tau = \pi\,{{\rm d}\over{\rm d}\ln\tau} \,,
\end{equation}
we find that
\begin{equation}\label{renSee}
   \Delta R_{\rm ren} = {1\over\beta_0}\,
   \int\limits_0^\infty\!{\rm d}\tau\,W_{e^+ e^-}(\tau)\,
   {\cal O}[-\delta_\tau]\,\delta(\ln\tau-\ln\tau_L) \nonumber\\
   = {1\over\beta_0}\,{\cal O}[\delta_\tau]\,\Big(
   \tau\,W_{e^+ e^-}(\tau) \Big)\bigg|_{\tau=\tau_L} \,,
\end{equation}
where $\tau_L=\Lambda_V^2/s$ is the position of the Landau pole, and
${\cal O}[\delta_\tau]$ denotes the differential operator
\begin{equation}
   {\cal O}[\delta_\tau] = 1 - {\delta_\tau^2\over 3!}
   + {\delta_\tau^4\over 5!} \mp \dots
   = {\sin\delta_\tau\over\delta_\tau} \,.
\end{equation}
To proceed, we use the relations
\begin{eqnarray}\label{Ooprel}
   {\cal O}[\delta_\tau]\,\tau^k &=& {\sin(k\pi)\over k\pi}\,
    \tau^k \,, \nonumber\\
   {\cal O}[\delta_\tau]\,\tau^k\,\ln^n\!\tau
   &=& \bigg( {{\rm d}\over{\rm d}k} \bigg)^n\,
    {\sin(k\pi)\over k\pi}\,\tau^k \,,
\end{eqnarray}
which imply that integer powers of $\tau$ in the distribution
function do not contribute to $\Delta R_{\rm ren}$ Given the
asymptotic behaviour shown in (\ref{asy}), we see that the leading
contribution to the renormalon ambiguity comes from the term
proportional to $\tau^2\ln\tau$. The result is
\begin{equation}\label{Reeren}
   \Delta R_{\rm ren} = -{4 C_F\over 3\beta_0}\,
   \bigg( {\Lambda_V^2\over s} \bigg)^3 + O(1/s^4) \,,
\end{equation}
in accordance with the fact that the nearest IR renormalon pole
in the Borel transform corresponding to the quantity $R(s)$ is
located at $u=3$ \cite{Broa,Bene}:\footnote{However, it is known that
a renormalon singularity at $u=2$ appears when one goes beyond the
large-$\beta_0$ approximation \protect\cite{reno5,Bene}.}
\begin{equation}\label{SRu}
   \widehat S_R(u) = {32 C_F\over\pi}\,{\sin(\pi u)\over u(2-u)}\,
   \sum_{k=2}^\infty\,{(-1)^k\,k\over\big[ k^2-(1-u)^2\big]^2}
   = -{4 C_F\over 3}\,{1\over 3-u} + \dots \,.
\end{equation}
The ellipses represent terms that are regular at $u=3$.

At this point a word of caution related to the interpretation of the
integral in (\ref{Srep}) is in order. In the case of the linear
integral representation (\ref{resum}), one can interpret the
integration variable $\tau$ as a physical scale parameter and
introduce a factorization point $\tau=\lambda^2/M^2$ to separate
short- and long-distance contributions. If one would proceed in the
same way in the present case, one would conclude that the
long-distance contribution to $R(s)$ is of order $\lambda^4/s^2$.
Such an interpretation may be misleading, however. In non-linear
representations there is no physical significance to the integration
variable; for instance, it is possible to obtain different non-linear
representations by changing variables or integrating by parts. An
example will be given in the following section. In the case of
(\ref{Srep}), one could try to obtain a linear representation by
repeating the steps used to derive the result for the renormalon
ambiguity given above. If $W_{e^+ e^-}(\tau)$ was an analytic
function, one would conclude that $(1/\tau)\,{\cal O}[\delta_\tau]\,
[\tau\,W_{e^+ e^-}(\tau)]$ is the distribution function for $R(s)$ in
the linear representation. Since higher derivatives of $W_{e^+
e^-}(\tau)$ are not continuous at $\tau=1$, however, one cannot
perform these steps. The resulting ``distribution function'' would
contain an infinite number of $\delta$-function type singularities at
$\tau=1$. In other words, for the physical correlation function
$R(s)$ a linear representation of the form (\ref{Srep}) does not
exist. As we will prove in Sect.~\ref{sec:add}, the same is true for
all quantities defined in the physical region.

Let us now discuss an alternative method to resum renormalon chains
for $R_{e^+ e^-}$. It is based on a representation of the function
$R(s)$ as an integral in the complex plane. Using that in the real
world the discontinuities of $D(-s)$ are located on the positive real
$s$-axis, and that $\mbox{Im}\,\Pi(0)=0$, one obtains
\begin{equation}\label{circle}
   R(s) = 4\pi\,\mbox{Im}\,\Pi(-s+i\epsilon)
   = {1\over\pi}\,\mbox{Im}\,\int\limits_0^s {{\rm d}s'\over s'}\,
   D(-s'+i\epsilon) = {1\over 2\pi i}\oint\limits_{|s'|=s}\!
   {{\rm d}s'\over s'}\,D(-s') \,.
\end{equation}
Inserting the expression (\ref{dQ2}) for the correlation function
and using the fact that
\begin{equation}
   a(-x\tau s) = {a(\tau s)\over 1 + a(\tau s)\ln(-x)}
\end{equation}
for the one-loop running coupling constant, we find
\begin{eqnarray}\label{Srepc}
   R_{\rm res}^{(2)}(s) &=& 1 + \int\limits_0^\infty\!{\rm d}\tau\,
    \widehat w_D(\tau)\,{1\over 2\pi i}
    \oint\limits_{|x|=1}\!{{\rm d}x\over x}\,
    {\alpha_s(-x\tau e^C s)\over 4\pi} \nonumber\\
   &=& 1 + {1\over\beta_0}\,\int\limits_0^\infty\!
    {\rm d}\tau\,W_{e^+ e^-}(\tau)\,F_2\big(a(\tau s)\big) \,,
\end{eqnarray}
where
\begin{equation}\label{Fcfun}
   F_2(a) = {1\over 2\pi i} \oint\limits_{|x|=1}\!
   {{\rm d}x\over x}\,{a\over 1 + a\ln(-x)}
   = {1\over\pi}\arctan(\pi a) \,.
\end{equation}
In Fig.~\ref{fig:3} the function $F_2(a)$ is compared to $F_1(a)$.
It is surprising that the resummations (\ref{Srep}) and (\ref{Srepc})
involve the same distribution function $W_{e^+ e^-}(\tau)$ but
different functions of the coupling constant. Clearly, both results
would have to be equivalent if the perturbative approximation for the
function $D(-s)$ would satisfy the analyticity properties of the
physical correlation function. However, it does not. Consider, for
example, the ``one-loop'' expression $D(-s) = 1+\alpha_s(-s)/\pi$.
The cut on the positive $s$-axis starts at $s=0$ (not $s>0$), and in
addition there is the Landau pole on the negative $s$-axis. The same
applies for the resummed series in (\ref{dQ2}); however, in this case
the Landau pole of the running coupling constant leads to a cut on
the negative $s$-axis. It is not difficult to show that these two
defects are responsible for the difference between the two
resummations. Note that the effective coupling constants $F_1(a)$ and
$F_2(a)$ are identical for $a>0$, implying that (\ref{Srep}) and
(\ref{Srepc}) correspond to the resummation of one and the same
perturbative series. The two functions differ only for negative
values of $a$, i.e.\ below the Landau pole. The difference between
the two resummations is thus a nonperturbative effect. We find
\begin{equation}
   R_{\rm res}^{(2)}(s) - R_{\rm res}^{(1)}(s) = -{1\over\beta_0}\,
   \int\limits_0^{\tau_L}\!{\rm d}\tau\,W_{e^+ e^-}(\tau)
   = -{3 C_F\over\beta_0}\,\bigg( {\Lambda_V^2\over s}
   \bigg)^2 + O(1/s^3) \,,
\end{equation}
where $\tau_L=\Lambda_V^2/s$ is the position of the Landau pole. Note
that this difference is parametrically larger than the IR renormalon
ambiguity of the resummed series given in (\ref{Reeren}). Indeed, the
effect is not related to renormalons; both resummations correspond to
the same perturbative series and thus to the same Borel transform. We
will come back to an analysis of this new type of ambiguity in
Sect.~\ref{sec:add}.

\section{Decay-rate ratio $R_\tau$}
\label{sec:4}

As a second example we consider the resummation of renormalon chains
for the perturbative series corresponding to the ratio
\begin{equation}
   R_\tau = {\Gamma(\tau\to\nu_\tau + \mbox{hadrons})
    \over\Gamma(\tau\to\nu_\tau\,e\,\bar\nu_e)}
   \equiv 3\,T(m_\tau^2) \,.
\end{equation}
For simplicity we neglect quark mass effects and small electroweak
corrections, which are however known \cite{Sirl,BraLi}. Our goal is
to understand in physical terms the low value of the BLM scale given
in the introduction, $\mu_{\rm BLM}^\tau\simeq 0.32\,m_\tau$ (in the
$\overline{\rm MS}$ scheme), and to investigate the effect of
higher-order corrections by performing the resummation of renormalon
chains. As in the previous case, $R_\tau$ can be expressed in terms
of the imaginary part of the correlator $\Pi(-s)$ in the physical
region $s>0$, or as a contour integral over the $D$-function in the
complex plane. We will again find that the result of the resummation
depends on which representation one chooses.

The first method is to relate $R_\tau$ to an integral over the
function $R(s)$ introduced in (\ref{SRdef}). For fixed neutrino
energy $E_\nu$, the invariant mass $s_{\rm had}$ of the hadronic
final state in the decay $\tau\to\nu_\tau+\mbox{hadrons}$ is given by
\begin{equation}
   s_{\rm had} = (p_\tau - p_\nu)^2 = m_\tau^2\,x \,,
   \qquad x = 1 - {2 E_\nu\over m_\tau} \in [0,1] \,.
\end{equation}
In the limit where one neglects the light quark masses, and as long
as one is interested only in terms of order $\beta_0^{n-1}\alpha_s^n$
in the perturbative series, the hadronic dynamics is described by the
function $R(s_{\rm had})$ discussed in the previous section. One
obtains \cite{Rtau1,Rtau2}
\begin{equation}\label{Stau}
   T(m_\tau^2) = 2\int\limits_0^1\!{\rm d}x\,
   (1 - 3 x^2 + 2 x^3)\,R(x m_\tau^2) \,,
\end{equation}
where the function $2(1-3 x^2+2 x^3)$ is the normalized invariant
mass spectrum at tree level. This expression allows us to understand
the low value of the BLM scale. We expect that
\begin{equation}
   \ln{(\mu_{\rm BLM}^\tau)^2\over m_\tau^2}
   = \ln{(\mu_{\rm BLM}^{e^+ e^-})^2\over s}
   + \langle\,\ln x\,\rangle \,,
\end{equation}
where
\begin{equation}
   \langle\,\ln x\,\rangle = 2\int\limits_0^1\!{\rm d}x\,
   (1 - 3 x^2 + 2 x^3)\,\ln x = - {19\over 12}
\end{equation}
is the average of $\ln x$ over the invariant mass spectrum. Hence,
the BLM scales corresponding to $R_{e^+ e^-}(m_\tau^2)$ and $R_\tau$
should obey the relation [see (\ref{muee})]
\begin{equation}
   \mu_{\rm BLM}^\tau = e^{-19/24}\,\mu_{\rm BLM}^{e^+ e^-}
   \simeq 0.738\,e^{C/2}\,m_\tau \,.
\end{equation}
We will see below that this is indeed the exact result. The fact that
the BLM scale for $R_\tau$ is lower than $m_\tau$ simply reflects the
distribution of the hadronic mass in the decay.

\begin{figure}[htb]
   \vspace{0.5cm}
   \epsfysize=6cm
   \centerline{\epsffile{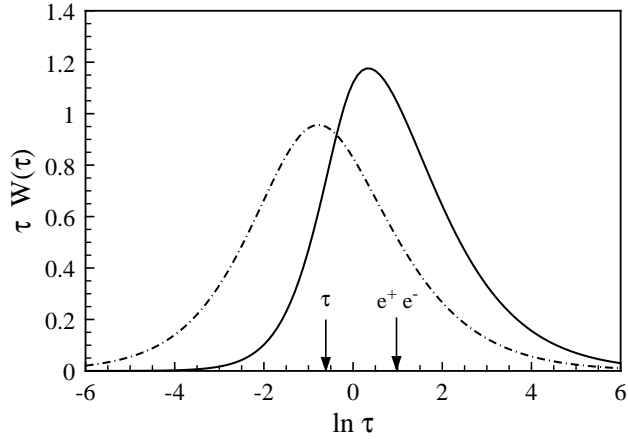}}
   \centerline{\parbox{13cm}{\caption{\label{fig:4}
The distribution functions $\tau\,W_{e^+ e^-}(\tau)$ (solid line) and
$\tau\,W_\tau(\tau)$ (dash-dotted line) as a function of $\ln\tau$.
The arrows indicate the average values of $\ln\tau$, which determine
the BLM scales.}}}
\end{figure}

Let us now proceed to construct the resummation of renormalon chains
for $R_\tau$. Substituting either (\ref{Srep}) or (\ref{Srepc}) into
(\ref{Stau}) and rescaling the integration variable
($\tau\to\tau/x$), we immediately obtain two different
representations for the resummed series. They are ($n=1$ or 2)
\begin{equation}\label{Sreptau}
   T_{\rm res}^{(n)}(m_\tau^2) = 1 + {1\over\beta_0}\,
   \int\limits_0^\infty\!{\rm d}\tau\,W_\tau(\tau)\,
   F_n\big(a(\tau m_\tau^2)\big) \,,
\end{equation}
with $F_n(a)$ as given in (\ref{Fdef}) and (\ref{Fcfun}),
respectively. The new distribution function is given by
\begin{equation}\label{wtaufun}
   W_\tau(\tau) = 2\int\limits_0^1{{\rm d}x\over x}\,
   (1 - 3 x^2 + 2 x^3)\,\widehat w_D\bigg( {\tau\over x} \bigg) \,.
\end{equation}
It is possible to perform this integral explicitly, with the result
that
\begin{eqnarray}
   W_\tau(\tau) &=& 16 C_F\,\Bigg\{ 4
    - {73\over 12}\,\tau - {23\over 24}\,\tau^2
    - {259\over 432}\,\tau^3 - 2 L_3(-\tau) - 3\zeta(3) \nonumber\\
   &&\qquad\mbox{}+ \bigg( {17\over 6}\,\tau + {\tau^2\over 3}
    + L_2(-\tau) \bigg)\,\ln\tau + \bigg( {3\over 4}\,\tau^2
    + {\tau^3\over 6} \bigg)\,\ln^2\!\tau \nonumber\\
   &&\qquad\mbox{}- \bigg( {11\over 6} + 3\tau
    + {3\over 2}\,\tau^2 + {\tau^3\over 3} \bigg)\,\Big[
    \ln\tau\,\ln(1+\tau) + L_2(-\tau) \Big] \Bigg\} \nonumber
\end{eqnarray}
for $\tau<1$, and
\begin{eqnarray}
   W_\tau(\tau) &=& 16 C_F\,\Bigg\{ - {575\over 216}
    + {37\over 48\tau} - {17\over 12}\,\tau
    - {\tau^2\over 3} + 2 L_3(-\tau^{-1}) \nonumber\\
   &&\qquad\mbox{}- \bigg( {85\over 36} - {1\over 4\tau}
    + {4\over 3}\,\tau + {\tau^2\over 3}
    - L_2(-\tau^{-1}) \bigg)\,\ln\tau \nonumber\\
   &&\qquad\mbox{}+ \bigg( {11\over 6} + 3\tau
    + {3\over 2}\,\tau^2 + {\tau^3\over 3} \bigg)\,\Big[
    \ln\tau\,\ln(1+\tau^{-1}) - L_2(-\tau^{-1}) \Big] \Bigg\}
\end{eqnarray}
for $\tau>1$. Here $L_3(x)=\int_0^x\frac{{\rm d}y}{y}\,L_2(y)$
denotes the trilogarithm function. In Fig.~\ref{fig:4} we compare the
distribution functions appearing in the resummations (\ref{Srep}) and
(\ref{Sreptau}) for $R_{e^+ e^-}$ and $R_\tau$. Note that the area
under the two curves is the same. The distribution function for
$R_\tau$ is broader than that for $R_{e^+ e^-}$ and shifted to lower
scales, in accordance with the physical argument presented above.

With the function $W_\tau(\tau)$ we compute the integrals
[cf.~(\ref{Deltadef})]
\begin{eqnarray}
   N &=& {3\over 4}\,C_F = 1 \,, \nonumber\\
   \langle \ln\tau \rangle &=& 4\zeta(3) - {65\over 12}
    \simeq -0.608 \,, \nonumber\\
   \langle \ln^2\!\tau \rangle &=& {2435\over 72}
    - {74\over 3}\,\zeta(3) \simeq 4.169 \,.
\end{eqnarray}
For the BLM scale and the parameter $\Delta_\tau$, which determines
the leading correction to the BLM scheme in (\ref{BLMrel}), we obtain
\begin{eqnarray}
   {\mu_{\rm BLM}^\tau\over m_\tau} &\simeq& 0.738\,e^{C/2}
    ~\stackrel{\overline{\rm MS}}{\to}~ 0.321 \,, \nonumber\\
   \Delta_\tau &=& \langle \ln^2\!\tau \rangle
    - \langle \ln\tau \rangle^2 - {\pi^2\over 3} \simeq 0.509 \,.
\end{eqnarray}

The asymptotic behaviour of the distribution function for $\tau\to 0$
is
\begin{equation}\label{asytau}
   W_\tau(\tau) = 16 C_F\,\bigg\{ 4 - 3\zeta(3) - {9\over 4}\,\tau
   + \bigg( {4\over 3} - {3\over 2}\,\ln\tau + {3\over 4}\,
   \ln^2\!\tau \bigg)\,\tau^2 + O(\tau^3) \bigg\} \,.
\end{equation}
Using (\ref{Ooprel}) we find that the renormalon ambiguity in the
value of the resummed series is given by
\begin{equation}\label{Tren}
   \Delta T_{\rm ren} = {8 C_F\over\beta_0}\,\bigg(
   \ln{m_\tau^2\over\Lambda_V^2} + {4\over 3} \bigg)
   \bigg( {\Lambda_V^2\over m_\tau^2} \bigg)^3 +O(1/m_\tau^8) \,,
\end{equation}
corresponding to an IR renormalon singularity located at $u=3$.
This can also be seen from the Borel transform of the series
$T(m_\tau^2)$, for which we obtain
\begin{eqnarray}\label{Stauu}
   \widehat S_T(u) &=& {384 C_F\over\pi}\,
    {\sin(\pi u)\over u(1-u)(2-u)(3-u)(4-u)}\,\sum_{k=2}^\infty\,
    {(-1)^k\,k\over\big[ k^2-(1-u)^2\big]^2} \nonumber\\
   &=& {8 C_F\over 3}\,\bigg\{ - {3\over(3-u)^2} + {4\over 3-u}
    + \dots \bigg\} \,.
\end{eqnarray}
The ellipses represent terms that are regular at $u=3$.

For later purposes it will be useful to rewrite the result
(\ref{Sreptau}) in another form. Instead of performing the
$x$-integration over the distribution function, one can perform it
over the function $F_n(a)$ and write ($n=1$ or 2)
\begin{equation}\label{taurep2}
   T_{\rm res}^{(n)}(m_\tau^2) = 1 + {1\over\beta_0}\,
   \int\limits_0^\infty\!{\rm d}\tau\,\Omega_\tau(\tau)\,
   G_n\big(a(\tau m_\tau^2)\big)
\end{equation}
with
\begin{equation}
   G_n(a) = 2\int\limits_0^1\!{\rm d}x\,(1 - 3 x^2 + 2 x^3)\,
   F_n\bigg( {a\over 1+a\ln x} \bigg) \,,
\end{equation}
and
\begin{equation}
   \Omega_\tau(\tau) = \widehat w_D(\tau) \,.
\end{equation}
Once again we have introduced a new notation for the distribution
function, since the form of the integral representation has changed.
Eqs.~(\ref{Sreptau}) and (\ref{taurep2}) provide an example of the
fact that the integration variable $\tau$ has no significance as a
physical scale if the integral representation is non-linear. Whereas
the contribution from the region $\tau<\lambda^2/m_\tau^2$ scales
like $\lambda^2/m_\tau^2$ in the case of (\ref{Sreptau}), it scales
like $\lambda^4/m_\tau^4$ in the case of (\ref{taurep2}).
Nevertheless, both representations are equivalent and lead to the
same numerical results.

\begin{figure}[htb]
   \vspace{0.5cm}
   \epsfysize=6cm
   \centerline{\epsffile{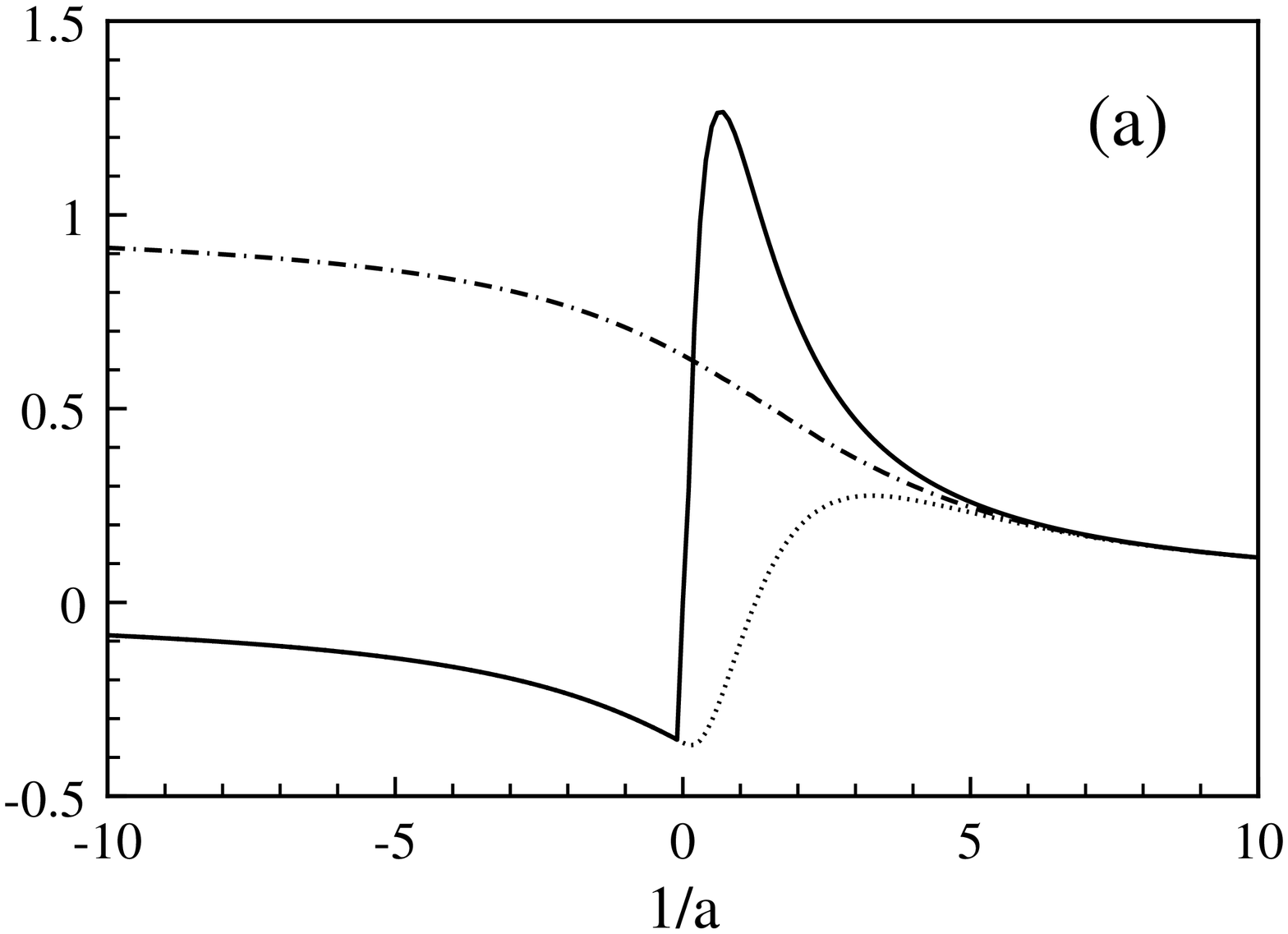}}
   \epsfysize=6cm
   \centerline{\epsffile{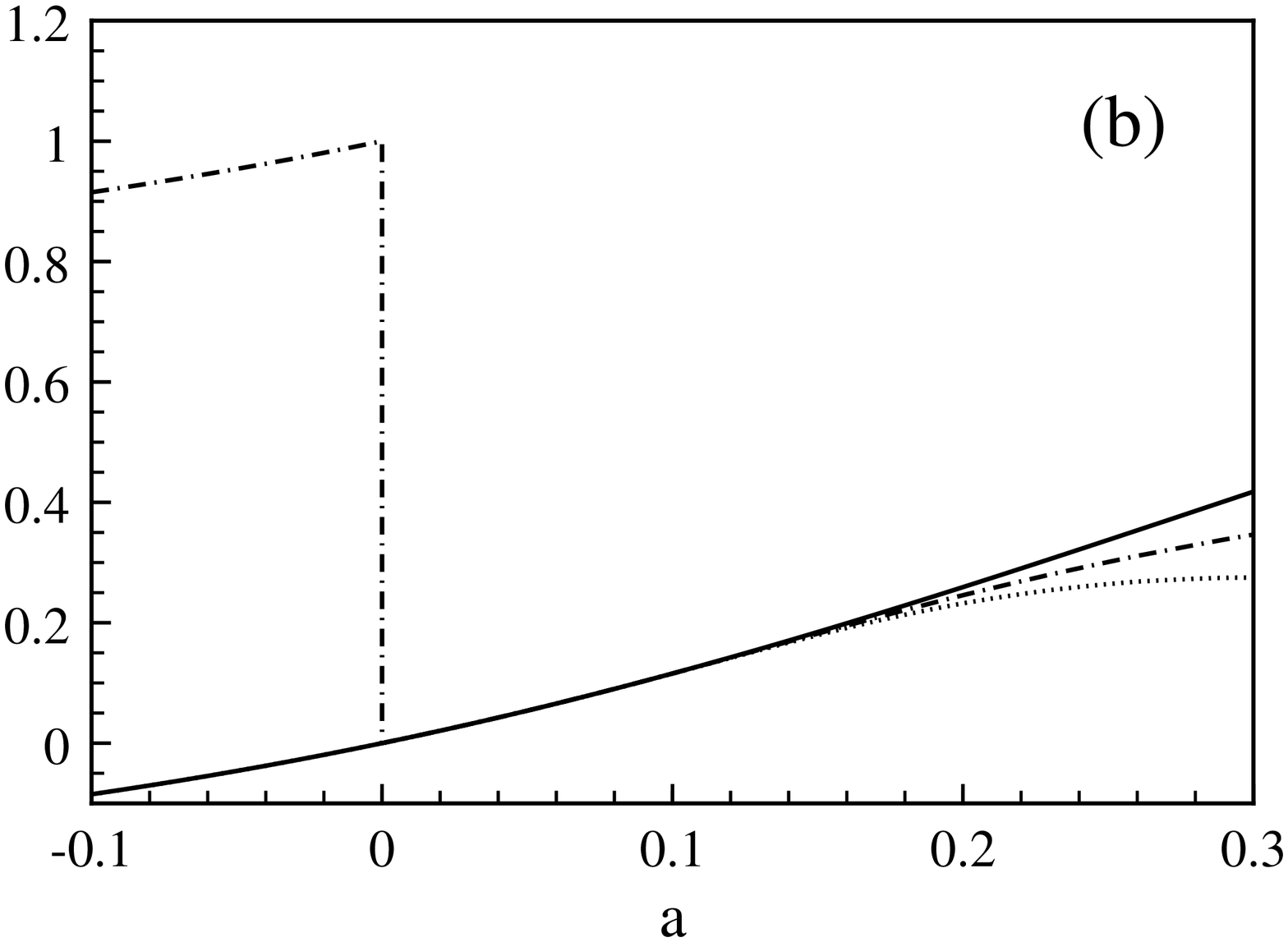}}
   \centerline{\parbox{13cm}{\caption{\label{fig:5}
Effective coupling constants $G_1(a)$ (dash-dotted line), $G_2(a)$
(dotted line) and $G_3(a)$ (solid line) as a function of (a) the
inverse coupling constant $1/a$, and (b) the coupling constant $a$.
For small positive values of $a$, the three functions have identical
Taylor expansions.}}}
\end{figure}

We will now discuss a third resummation method, which uses the
representation of $T(m_\tau^2)$ as a contour integral in the complex
plane. Since in the real world the discontinuities of the function
$D(-s)$ lie on the positive real $s$-axis and start at a value
$s_0>0$, one can integrate by parts in (\ref{Stau}) to obtain
\begin{eqnarray}\label{tcircle}
   T(m_\tau^2) &=& {1\over\pi}\,\mbox{Im}
    \int\limits_0^1{{\rm d}x\over x}\,
    (1 - 2 x + 2 x^3 - x^4)\,D(-x m_\tau^2+i\epsilon) \nonumber\\
   &=& {1\over 2\pi i}\,\oint\limits_{|x|=1}\!{{\rm d}x\over x}\,
    (1 - 2 x + 2 x^3 - x^4)\,D(-x m_\tau^2) \,.
\end{eqnarray}
Inserting the expression (\ref{dQ2}) for the $D$-function we find
\begin{equation}\label{taurep3}
   T_{\rm res}^{\rm circle}(m_\tau^2) = 1 + {1\over\beta_0}\,
   \int\limits_0^\infty\!{\rm d}\tau\,\Omega_\tau(\tau)\,
   G_3\big(a(\tau m_\tau^2)\big) \,,
\end{equation}
where
\begin{eqnarray}\label{G3fun}
   G_3(a) &=& {1\over 2\pi i}\,\oint\limits_{|x|=1}\!
    {{\rm d}x\over x}\,(1 - 2 x + 2 x^3 - x^4)\,
    {a\over 1 + a\ln(-x)} \nonumber\\
   &=& {8 a\over\pi}\,\int\limits_0^\pi\!{\rm d}\varphi\,
    {\sin 2\varphi - a\varphi \cos 2\varphi \over
     1 + a^2\varphi^2}\,\sin\varphi\,\cos^2{\varphi\over 2}
\end{eqnarray}
is one of the functions encountered in Ref.~\cite{Rt5}.
Eqs.~(\ref{taurep2}) and (\ref{taurep3}) provide three resummations
of the perturbative series for $T(m_\tau^2)$ that have the same
distribution function $\Omega_\tau(\tau)$, but different effective
coupling constants $G_n(a)$. These three functions are shown in
Fig.~\ref{fig:5}. For small positive values of $a$ they have
identical Taylor expansions, meaning that all three resummations
correspond to one and the same perturbative series. Outside the
radius of convergence the functions $G_n(a)$ are quite different,
however, leading to differences in the value of the resummed series
which have the form of nonperturbative power corrections. Again these
nonperturbative effects have nothing to do with IR renormalons. It is
not difficult to trace how the differences arise from the wrong
analytic properties of the perturbative expression for the function
$D(-s)$, which has Landau pole singularities on the negative $s$-axis
and a cut on the positive $s$-axis starting at $s=0$ (not $s>0$). We
find that
\begin{eqnarray}\label{Gadiffs}
   G_2(a) - G_1(a) &=& \cases{ -2 e^{-1/a} + 2 e^{-3/a} - e^{-4/a}
    &; $a>0$ \,, \cr
    -1 &; $a<0$ \,, \cr} \nonumber\\
   G_3(a) - G_2(a) &=& 4 (e^{-1/a} - e^{-3/a})\,\Theta(a) \,.
\end{eqnarray}
For the differences between the three resummations we obtain, using
(\ref{Sreptau}) and (\ref{Gadiffs}),
\begin{eqnarray}\label{T123}
   T_{\rm res}^{(2)}(m_\tau^2) - T_{\rm res}^{(1)}(m_\tau^2)
   &=& -{1\over\beta_0}\,\int\limits_0^{\tau_L}\!{\rm d}\tau\,
    W_\tau(\tau) = -{16 C_F\over\beta_0}\,
    \Big[ 4 - 3\zeta(3) \Big]\,{\Lambda_V^2\over m_\tau^2}
    + O(1/m_\tau^4) \,, \nonumber\\
   T_{\rm res}^{\rm circle}(m_\tau^2) - T_{\rm res}^{(2)}(m_\tau^2)
   &=& {4\over\beta_0}\,\int\limits_{\tau_L}^\infty\!
    {\rm d}\tau\,\Omega_\tau(\tau)\,\bigg( {\tau_L\over\tau}
    - {\tau_L^3\over\tau^3} \bigg) \nonumber\\
   &=& {32 C_F\over\beta_0}\,\Big[ 4 - 3\zeta(3) \Big]\,
    {\Lambda_V^2\over m_\tau^2} + O(1/m_\tau^4) \,,
\end{eqnarray}
where $\tau_L=\Lambda_V^2/m_\tau^2$ is the position of the Landau
pole. As in the case of $R_{e^+ e^-}$, the differences are
parametrically larger than the renormalon ambiguity given in
(\ref{Tren}). In the present case they are of order $1/m_\tau^2$ and
thus numerically quite significant; we note that the terms shown on
the right-hand side in (\ref{T123}) equal $-6.3\%$ and $12.6\%$. A
new feature of the present case is that the effective coupling
constants $G_n(a)$ differ not only for $a<0$, but also in the region
above the Landau pole (although by exponentially small terms). This
is what makes the various resummations differ by terms of order
$1/m_\tau^2$. We will argue in the next section that such differences
can be avoided. There is a physical argument why one should prefer
one form of the resummation, $T_{\rm res}^{\rm circle}$, over the
others. However, as in the case of $R_{e^+ e^-}$ there will remain
ambiguities related to the treatment of the region below the Landau
pole (for $a<0$), which are of order $1/m_\tau^4$. These effects are
still larger than the IR renormalon ambiguity.

\section{Cutoff regularization of the Borel integral}
\label{sec:add}

The goal of this section is to understand better the origin of the
ambiguities encountered in the resummation of renormalon chains for
$R_{e^+ e^-}$ and $R_\tau$. We recall that a surprising feature of
these nonperturbative effects is that they are not related to IR
renormalon singularities in the Borel plane; the differences between
the various resummations are parametrically larger than the ambiguity
resulting from the nearest IR renormalon. Therefore, it is tempting
to conjecture that these differences reflect a failure of our
resummation procedure, which could be avoided by defining the
resummation in terms of the Borel integral (\ref{Laplace}), in which
case only the usual IR renormalon ambiguities appear. This conjecture
may be wrong, however. It has been noted already by 't Hooft that
most likely the Borel integral is ill-defined because of strong
singularities of the Borel transform at infinity \cite{tHof}. The
presence of resonances and multi-particle thresholds in the physical
region is in conflict with Borel summation. It is argued in
Ref.~\cite{tHof} that, as a consequence, the Borel transform
$\widehat S(u)$ must diverge at $u\to\infty$ stronger than any
exponential of $u$. Discussions of the implications of this
observation can be found in Refs.~\cite{BenOPE,ShifOPE}. We note that
the strong singular behaviour for large values of $u$ is not seen if
the Borel transform is calculated using dimensional regularization.
The situation is similar to power-divergent Feynman integrals of the
type $\int{\rm d}^d k\,(k^2)^n$, which vanish by definition in
dimensional regularization. Let us therefore study the singularities
of the Borel transform in a more careful way.

\subsection{Cutoff regularization in the euclidean region}

The proper way to analyze long-distance effects is by introducing an
IR cutoff $\lambda$, which eliminates the low-momentum contributions
from perturbative calculations. Let us investigate the connection
between this procedure and the Borel integral in the euclidean
region, where the OPE provides a consistent framework to separate
short- and long-distance contributions \cite{Wils}. As shown in
(\ref{Swrela}), the Borel transform $\widehat S(u)$ can be defined in
terms of an integral over a distribution function $\widehat w(\tau)$,
which has a concrete physical interpretation. This integral has a
finite radius of convergence; it is well defined if
$u\in\,]\!-\!j,k\,[$, where $j$ and $k$ denote the positions of the
nearest UV and IR renormalon singularities. The standard way to
proceed is to define $\widehat S(u)$ outside this interval by
analytic continuation, in which case the divergent behaviour of the
integral for large or small values of $\tau$ is reflected in the
appearance of renormalon singularities. The Borel integral
(\ref{Laplace}) has an ambiguity of order $(\Lambda_V^2/M^2)^k$ due
to the nearest IR renormalon. It is usually argued that the
appearance of renormalon ambiguities signals the size of
nonperturbative corrections, which have to be added to perturbative
calculations in QCD \cite{reno2}--\cite{Muel}. In order to understand
the link between the Borel integral and the OPE, let us define a
regularized Borel transform as
\begin{equation}
   \widehat S_{\rm reg}(u,\tau_0) = \int\limits_{\tau_0}^\infty\!
   {\rm d}\tau\,\widehat w(\tau)\,\tau^{-u} \,;\quad
   \tau_0 = {\lambda^2\over M^2} \,.
\end{equation}
The dependence of $\tau_0$ on $M$ is dictated by the fact that,
according to (\ref{resum}), the product $\sqrt{\tau} M$ has the
interpretation of a physical scale. The introduction of this cutoff
eliminates all IR renormalon singularities.\footnote{A similar
regularization can be applied to eliminate the UV renormalon
singularities.}
In particular, if the asymptotic behaviour of the distribution
function for $\tau\to 0$ is as shown in (\ref{wtauexp}), the
behaviour of the regularized Borel transform in the vicinity of $u=k$
is
\begin{equation}
   \widehat S_{\rm reg}(u,\tau_0) = {w_0\over u-k}\,\Big(
   \tau_0^{k-u} - 1 \Big) + \dots = w_0\,\ln{M^2\over\lambda^2}
   + \dots
\end{equation}
instead of $\widehat S(u)=-w_0/(u-k)+\dots$, which is obtained
without using a cutoff. The regularization eliminates all
singularities for $u>0$, but at the same time it leads to a strong
exponential growth of the Borel transform for $u\to\infty$:
\begin{equation}
   \widehat S_{\rm reg}(u,\tau_0) \sim \exp\bigg(
   u\,\ln{1\over\tau_0} \bigg) = \exp\bigg(
   u\,\ln{M^2\over\lambda^2} \bigg) \,.
\end{equation}
This growth becomes arbitrarily strong as one tries to eliminate the
IR cutoff, in accordance with 't Hooft's conjecture.

The regularized form of the Borel integral reads
\begin{equation}\label{revers}
   S_{\rm Borel}^{\rm reg}(M^2,\tau_0) = 1 + {1\over\beta_0}\,
   \int\limits_0^\infty\!{\rm d}u\,\widehat S_{\rm reg}(u,\tau_0)\,
   \bigg( {\Lambda_V^2\over M^2} \bigg)^u
   = 1 + {1\over\beta_0}\,\int\limits_0^\infty\!{\rm d}u
   \int\limits_{\tau_0}^\infty\!{\rm d}\tau\,
   \widehat w(\tau)\,\bigg( {\tau M^2\over\Lambda_V^2}
   \bigg)^{-u} \,.
\end{equation}
The integral over $u$ is convergent if $\tau_0>\Lambda_V^2/M^2$, or
equivalently if $\lambda>\Lambda_V$, which we have to assume for any
sensible IR regulator. It is then allowed to interchange the order of
integration and perform the integral over $u$ first, which gives
\begin{equation}\label{uint}
   \int\limits_0^\infty\!{\rm d}u\,\bigg( {\tau M^2\over\Lambda_V^2}
   \bigg)^{-u} = {1\over\ln(\tau M^2/\Lambda_V^2)}
   = {\beta_0\,\alpha_s(\tau e^C M^2)\over 4\pi} \,.
\end{equation}
This leads to
\begin{equation}\label{Borelreg}
   S_{\rm Borel}^{\rm reg}(M^2,\tau_0) = 1
   + \int\limits_{\tau_0}^\infty\!{\rm d}\tau\,
   \widehat w(\tau)\,{\alpha_s(\tau e^C M^2)\over 4\pi}
   = 1 + S_{\rm sd}(M^2,\lambda^2)\Big|_{\lambda^2=\tau_0 M^2} \,,
\end{equation}
which is nothing but the short-distance contribution to resummed
series defined in (\ref{separ}). The standard form of the Borel
integral is recovered if one sets $\tau_0=0$ and defines the
right-hand side of (\ref{uint}) to be the analytic continuation of
the $u$-integral in the region below the Landau pole, i.e.\ for
$\tau<\Lambda_V^2/M^2$. Then the integration over the Landau pole
leads to an ambiguity of order $(\Lambda_V^2/M^2)^k$ in
(\ref{Borelreg}), which is the usual IR renormalon ambiguity.
However, the more physical way to proceed is to keep the IR cutoff.
In the regularized version of the Borel integral it is the way in
which the factorization scale appears that determines the size of
long-distance contributions, which have to be added to the above
result in order to cancel the dependence on $\lambda$. From the
renormalization-group analysis in Sect.~\ref{sec:2}, it follows that
this long-distance contribution has the same $M$-dependence as the
renormalon ambiguity [compare (\ref{Dren}) with (\ref{lamRGE})].
Hence, the IR behaviour in the cutoff version of the Borel integral
is in one-to-one correspondence with the appearance of renormalon
singularities in the Borel plane. It is this fact which, a
posteriori, gives a physical significance to these ambiguities, and
which establishes the connection between the Borel integral and the
OPE in the euclidean region.

Let us now investigate what changes if the quantity of interest is
defined in the physical region. The crucial difference is that in
this case the integral representation of the Borel transform shown in
(\ref{Swrela}) does not exist. We will now prove this statement,
which was already mentioned in the discussion of $R_{e^+ e^-}$ in
Sect.~\ref{sec:3}. Consider the following alternative integral
representation for the Borel transform $\widehat S(u)$ of a quantity
$S(M^2)$:
\begin{equation}\label{BBrepr}
   \widehat S(u) = {\sin\pi u\over\pi u}\,
   \int\limits_0^\infty\!{\rm d}\tau\,W(\tau)\,\tau^{-u} \,.
\end{equation}
Beneke and Braun have shown that this representation always exists
\cite{BBnew}. Let us assume that also the representation
(\ref{Swrela}) exists. Then there is a simple relation between the
functions $\widehat w(\tau)$ and $W(\tau)$. Rewriting the
sin-function in terms of $\Gamma$-functions, we find that
\begin{eqnarray}
   \int\limits_0^\infty\!{\rm d}\tau\,W(\tau)\,\tau^{-u}
   &=& \Gamma(1+u)\,\Gamma(1-u)\,\int\limits_0^\infty\!
    {\rm d}\tau\,\widehat w(\tau)\,\tau^{-u} \nonumber\\
   &=& \int\limits_0^\infty\!{\rm d}\tau\,\widehat w(\tau)\,
     \tau^{-u} \int\limits_0^1\!{\rm d}x\,x^u (1-x)^{-u} \,,
\end{eqnarray}
and introducing new variables $z=x/(1-x)$ and $t=\tau/z$ we
obtain
\begin{equation}
   \int\limits_0^\infty\!{\rm d}\tau\,W(\tau)\,\tau^{-u}
   = \int\limits_0^\infty\!{\rm d}t
   \int\limits_0^\infty\!{\rm d}z\,{z\over(1+z)^2}\,
   \widehat w(z t)\,t^{-u} \,.
\end{equation}
{}From this equation we read off the relation ($\sigma=z t$)
\begin{equation}\label{Wtwtau}
   W(\tau) = \int\limits_0^\infty\!{\rm d}\sigma\,
   {\sigma\,\widehat w(\sigma)\over(\sigma+\tau)^2}
   = \int\limits_0^\infty\!{\rm d}\sigma\,
   {1\over\sigma+\tau}\,{{\rm d}\over{\rm d}\sigma}\,\Big[
   \sigma\,\widehat w(\sigma) \Big] \,,
\end{equation}
which establishes the connection between the two
functions.\footnote{In the last step we have assumed that the
quantity $S(M^2)$ is at most logarithmically UV divergent, so that
the distribution function $\widehat w(\tau)$ vanishes for
$\tau=\infty$.}
Provided a function $\widehat w(\tau)$ satisfying the integral
representation (\ref{Swrela}) exists, it follows that the function
$W(\tau)$ is analytic in the complex $\tau$-plane with a branch cut
along the negative axis. We can now use the observation of Beneke and
Braun that the function $W(\tau)$ can be calculated by performing
loop integrals with a finite gluon mass $m_g^2=\tau M^2$
\cite{BBnew}. In the euclidean region these integrals have
discontinuities only for unphysical values $m_g^2<0$, and the
function $W(\tau)$ has indeed the analyticity properties following
from (\ref{Wtwtau}). However, in the physical region there is the
possibility of real gluon emission if $m_g^2<M^2$, and hence the
function $W(\tau)$ must contain a contribution proportional to
$\Theta(1-\tau)$, which is in conflict with (\ref{Wtwtau}).
Therefore, a function $\widehat w(\tau)$ satisfying the above
integral relation does not exist in the physical region.

It follows then that a ``linear'' integral representation of the type
(\ref{resum}) does not exist, too. One cannot avoid to use
``non-linear'' representations, such as the ones encountered in
Sects.~\ref{sec:3} and \ref{sec:4}. Because the form of the
non-linear representation is not unique, however, it is no longer
obvious how to introduce a factorization scale in order to separate
short- and long-distance contributions. A possible way to proceed is
to define the resummation of renormalon chains in terms of the Borel
integral. However, as in the euclidean region one should check that
the resulting ambiguities from IR renormalons are in one-to-one
correspondence with the size of long-distance contributions. To this
end one should introduce an IR cutoff. A possibility which seems well
motivated to us is first to regulate the Borel transform in the
euclidean region, and then to perform the analytic continuation to
the physical region. We shall now discuss this proposal for the
quantities $R(s)$ and $T(m_\tau^2)$.

\subsection{Cutoff regularization for $R(s)$}

We start with the regularized form of the Borel integral for the
correlator $D(Q^2)$ in the euclidean region, which reads
\begin{equation}\label{Dreg}
   D_{\rm Borel}^{\rm reg}(Q^2,\tau_0) = 1
   + {1\over\beta_0}\,\int\limits_0^\infty\!{\rm d}u
   \int\limits_{\tau_0}^\infty\!{\rm d}\tau\,\widehat w_D(\tau)\,
   \bigg( {\tau Q^2\over\Lambda_V^2} \bigg)^{-u} \,.
\end{equation}
The integral over $u$ is convergent as long as
$\tau_0>\Lambda_V^2/Q^2$. Let us now consider the analytic
continuation of the above result in the complex plane, which is well
defined if $\tau_0>\Lambda_V^2/|Q^2|$. Outside a circle of radius
$\Lambda_V^2/\tau_0$ around the origin, the regularized function
$D_{\rm Borel}^{\rm reg}(Q^2,\tau_0)$ is analytic in the complex
$Q^2$-plane with a branch cut along the negative axis. Using the
above expression, we find that both methods of calculating the
function $R(s)$ -- taking the imaginary part of $\Pi(-s+i\epsilon)$,
or integrating $D(Q^2)$ around a circle of radius $s$ in the complex
plane -- lead to the same result:\footnote{Note that this is
equivalent to using a representation of the form
(\protect\ref{BBrepr}) for the Borel transform of $R(s)$, provided we
identify $W_{e^+ e^-}(\tau)=\widehat w_D(\tau)$ [see
(\protect\ref{WeewD})]. This relation is in accordance with the fact
that the Borel transform of $R(s)$ in (\protect\ref{SRu}) differs
from the Borel transform of $D(Q^2)$ in (\protect\ref{SDu}) by a
factor $\sin(\pi u)/(\pi u)$.}
\begin{equation}
   R_{\rm reg}(s,\tau_0) = 1 + {1\over\beta_0}\,
   \int\limits_0^\infty\!{\rm d}u\,{\sin(\pi u)\over\pi u}
   \int\limits_{\tau_0}^\infty\!{\rm d}\tau\,
   \widehat w_D(\tau)\,\bigg( {\tau s\over\Lambda_V^2}
   \bigg)^{-u} \,.
\end{equation}
As long as $\tau_0>\Lambda_V^2/s$, both integrals are convergent and
we can perform the integral over $u$ first. This gives
\begin{equation}\label{Finte}
   \int\limits_0^\infty\!{\rm d}u\,{\sin(\pi u)\over\pi u}\,
   \bigg( {\tau s\over\Lambda_V^2} \bigg)^{-u}
   = \int\limits_0^\infty\!{\rm d}u\,{\sin(\pi u)\over\pi u}\,
   e^{-u/a} = F(a) \,,
\end{equation}
where $a=1/\ln(\tau s/\Lambda_V^2)$ agrees with the coupling constant
$a(\tau s)$ defined in (\ref{adef}), and
\begin{equation}\label{Fill}
   F(a) = {1\over\pi}\,\mbox{arccot}\,\bigg( {1\over\pi a} \bigg)
   = {1\over\pi}\,\arctan(\pi a) \,;\quad \mbox{for}~a>0 \,.
\end{equation}
Note that $a>0$ if $\tau>\Lambda_V^2/s$. Setting now
$\tau_0=\lambda^2/s$ with $\lambda>\Lambda_V$, and writing the final
result as a function of $s$ and $\lambda^2$ (instead of $\tau_0$), we
obtain
\begin{equation}\label{Rreg}
   R_{\rm reg}(s,\lambda^2) = 1 + {1\over\beta_0}\,
   \int\limits_{\lambda^2/s}^\infty\!{\rm d}\tau\,
   W_{e^+ e^-}(\tau)\,F\big( a(\tau s) \big) \,.
\end{equation}
For $a>0$, the effective coupling constant $F(a)$ coincides with the
functions $F_1(a)$ and $F_2(a)$ in (\ref{Fdef}) and (\ref{Fcfun}).
However, if one wants to remove the cutoff one has to define $F(a)$
in the region $a<0$, i.e.\ below the Landau pole, where the integral
in (\ref{Finte}) is ill defined. Both $F_1(a)$ and $F_2(a)$ are
``natural'' continuations of $F(a)$; one is continuous as a function
of $1/a$, the other is continuous as a function of $a$. The fact that
the choice of the continuation below the Landau pole is not unique
leads to an ambiguity in the definition of the resummed series. In
the case of the physical correlator $R(s)$, we have seen in
Sect.~\ref{sec:3} that this ambiguity is parametrically larger than
the IR renormalon ambiguity.

In this context it is interesting to note that neither $R_{\rm
res}^{(1)}(s)$ nor $R_{\rm res}^{(2)}(s)$ coincide with the principle
value of the Borel integral. This is different from the situation
encountered in the euclidean region, where taking the limit
$\lambda\to 0$ in (\ref{Borelreg}) one reproduces the value of the
Borel integral. In the present case, however, Beneke and Braun have
shown that the principle value of the Borel integral is given by (see
the Appendix) \cite{BBnew}
\begin{equation}\label{RsBorel}
   R_{\rm Borel}(s) = 1 + {1\over\beta_0}\,
   \int\limits_0^\infty\!{\rm d}\tau\,W_{e^+ e^-}(\tau)\,
   F_1\big( a(\tau s) \big) + {1\over\beta_0}\,\mbox{Re}
   \int\limits_{-\tau_L}^0\!{\rm d}\tau\,
   W_{e^+ e^-}(\tau-i\epsilon) \,,
\end{equation}
where $\tau_L=\Lambda_V^2/s$ is the position of the Landau pole. The
imaginary part of the second integral determines the renormalon
ambiguity:
\begin{equation}
   \Delta R_{\rm ren} = {1\over\pi\beta_0}\,\mbox{Im}
   \int\limits_{-\tau_L}^0\!{\rm d}\tau\,
   W_{e^+ e^-}(\tau-i\epsilon) \,.
\end{equation}
Comparing (\ref{RsBorel}) with (\ref{Srep}) and (\ref{Srepc}), we
find that
\begin{eqnarray}\label{RBorcomp}
   R_{\rm Borel}(s) - R_{\rm res}^{(1)}(s)
   &=& {1\over\beta_0}\,\mbox{Re} \int\limits_{-\tau_L}^0\!
    {\rm d}\tau\,W_{e^+ e^-}(\tau-i\epsilon)
    = - {3 C_F\over\beta_0}\,\bigg( {\Lambda_V^2\over s} \bigg)^2
    + O(1/s^3) \,, \nonumber\\
   R_{\rm Borel}(s) - R_{\rm res}^{(2)}(s)
   &=& {1\over\beta_0}\,\mbox{Re} \int\limits_{-\tau_L}^{\tau_L}\!
    {\rm d}\tau\,W_{e^+ e^-}(\tau-i\epsilon) \nonumber\\
   &=& - {8 C_F\over 3\beta_0}\,\bigg( \ln{s\over\Lambda_V^2}
    + {11\over 6} \bigg) \bigg( {\Lambda_V^2\over s} \bigg)^3
    + O(1/s^4) \,.
\end{eqnarray}
Moreover, we obtain
\begin{equation}
   \Delta R_{\rm ren} = {4 C_F\over\beta_0}\,\bigg\{
   \tau_L - {3\over 2}\,\tau_L^2 + (1-\tau_L)^2\,\ln(1-\tau_L)
   \bigg\} \,,
\end{equation}
which, when expanded in powers of $\tau_L=\Lambda_V^2/s$, reproduces
(\ref{Reeren}). The ambiguities arising from the choice of the
continuation below the Landau pole lead to differences of order
$1/s^2$, which are parametrically larger than the IR renormalon
ambiguity.

It is not difficult to understand the size of these effects by
considering the regularized form of the Borel integral given in
(\ref{Rreg}). In fact, instead of taking the limit $\lambda\to 0$,
the
more physical way to proceed is to keep the IR cutoff in the
perturbative calculation, and to add to the regularized result a
scale-dependent nonperturbative contribution, so that the dependence
on the cutoff cancels in the sum. From the renormalization-group
equation
\begin{eqnarray}
   \lambda^2\,{{\rm d}\over{\rm d}\lambda^2}\,
   R_{\rm reg}(s,\lambda^2) &=& - {1\over\beta_0}\,
    F\big( a(\lambda^2) \big)\,{\lambda^2\over s}\,
    W_{e^+ e^-}(\lambda^2/s) \nonumber\\
   &=& - {3 C_F\over\beta_0}\,F\big( a(\lambda^2) \big)\,
    \bigg( {\lambda^2\over s} \bigg)^2 + O(1/s^3) \,,
\end{eqnarray}
it follows that this nonperturbative contribution must scale like
$1/s^2$. At this point we have to add a word of caution, however.
Contrary to the euclidean region, one has to be careful when
interpreting $\lambda$ as a physical scale parameter. Although the
introduction of the IR cutoff certainly removes all soft
contributions, we cannot exclude that it actually removes too much,
i.e.\ contributions which may come from short distances as well.
Unless one is able to construct a cutoff regularization for which the
nonperturbative contribution scales like $1/s^3$, however, one
should be conservative and assume that there are indeed long-distance
contributions of order $1/s^2$.

\subsection{Cutoff regularization for $T(m_\tau^2)$}

Let us now turn to the more complicated case of the function
$T(m_\tau^2)$. If we work with the representation of this quantity as
an integral over $R(s)$ as shown in (\ref{Stau}) and insert the
regularized expression given in (\ref{Rreg}), we find
\begin{equation}\label{Tres1}
   T_{\rm reg}(m_\tau^2,\lambda^2) = 1 + {2\over\beta_0}\,
   \int\limits_0^1\!{\rm d}x\,(1-3 x^2+2 x^3)\,
   \int\limits_0^\infty\!{\rm d}u\,{\sin(\pi u)\over\pi u}\,
   \int\limits_{\lambda^2/x m_\tau^2}^\infty\!{\rm d}\tau\,
   \widehat w_D(\tau)\,\bigg( {\tau x m_\tau^2\over\Lambda_V^2}
   \bigg)^{-u} \,.
\end{equation}
Note that in order for the $u$-integral to be convergent we have to
introduce an $x$-dependent cutoff $\tau_0=\lambda^2/(x m_\tau^2)
=\lambda^2/s_{\rm had}$ with $\lambda>\Lambda_V$, where $s_{\rm had}$
is the invariant mass of the hadronic final state. It is thus
convenient to introduce a new variable $\tau'=\tau x$ and to perform
the integral over $x$ at fixed $\tau'$ to obtain
\begin{eqnarray}
   T_{\rm reg}(m_\tau^2,\lambda^2) &=& 1 + {1\over\beta_0}\,
    \int\limits_0^\infty\!{\rm d}u\,{\sin(\pi u)\over\pi u}\,
    \int\limits_{\lambda^2/m_\tau^2}^\infty\!{\rm d}\tau'\,
    W_\tau(\tau') \bigg( {\tau' m_\tau^2\over\Lambda_V^2}
    \bigg)^{-u} \nonumber\\
   &=& 1 + {1\over\beta_0}\,
    \int\limits_{\lambda^2/m_\tau^2}^\infty\!{\rm d}\tau\,
    W_\tau(\tau)\,F\big(a(\tau m_\tau^2)\big) \,,
\end{eqnarray}
where $W_\tau(\tau)$ is the distribution function defined in
(\ref{wtaufun}), and $F(a)$ has been given in (\ref{Fill}). The
freedom in the choice of the continuation to the region below the
Landau pole (where $a<0$) introduces an ambiguity of order
$1/m_\tau^2$, since $W_\tau(\tau)\to\mbox{const.}$ for $\tau\to 0$.
Two examples of such continuations are provided by the quantities
$T_{\rm res}^{(1)}(m_\tau^2)$ and $T_{\rm res}^{(2)}(m_\tau^2)$ in
(\ref{Sreptau}), which indeed differ by terms of order $1/m_\tau^2$
[see (\ref{T123})]. If the IR cutoff is kept, the ambiguity of
order $1/m_\tau^2$ is reflected in the fact that
\begin{eqnarray}
   \lambda^2\,{{\rm d}\over{\rm d}\lambda^2}\,
   T_{\rm reg}(m_\tau^2,\lambda^2) &=& - {1\over\beta_0}\,
    F\big( a(\lambda^2) \big)\,{\lambda^2\over m_\tau^2}\,
    W_\tau(\lambda^2/m_\tau^2) \nonumber\\
   &=& - {16 C_F\over\beta_0}\,\Big[ 4 - 3\zeta(3) \Big]\,
    F\big( a(\lambda^2) \big)\,{\lambda^2\over m_\tau^2}
    + O(1/m_\tau^4) \,,
\end{eqnarray}
which means that one has to add a nonperturbative contribution of
order $1/m_\tau^2$ to cancel the dependence on the cutoff.

Let us now consider the representation of $T(m_\tau^2)$ as a contour
integral given in (\ref{tcircle}). Inserting there the regularized
expression for the function $D(Q^2)$ from (\ref{Dreg}), we find
\begin{equation}\label{Tres2}
   T_{\rm reg}^{\rm circle}(m_\tau^2,\lambda^2) = 1
   + {1\over\beta_0}\,{1\over 2\pi i}\oint\limits_{|x|=1}\!
   {{\rm d}x\over x}\,(1-2 x+2 x^3-x^4)\,
   \int\limits_0^\infty\!{\rm d}u\,
   \int\limits_{\tau_0}^\infty\!{\rm d}\tau\,\widehat w_D(\tau)\,
   \bigg( {\tau x m_\tau^2\over\Lambda_V^2} \bigg)^{-u} \,,
\end{equation}
where $\tau_0=\lambda^2/|x m_\tau^2|=\lambda^2/m_\tau^2$. This
means that the integral over $x$ can be performed without affecting
the value of the IR cutoff, which is determined by the radius of the
contour in the complex plane. The result is\footnote{The same result
is obtained if one performs first the $x$-integral in
(\protect\ref{Tres1}), which is convergent for $u<1$, and then
analytically continues the result to arbitrary values of $u$.
However, this procedure involves an integration over regions where
the Borel integral is ill defined.}
\begin{equation}
   T_{\rm reg}^{\rm circle}(m_\tau^2,\lambda^2) = 1
   + {1\over\beta_0}\,\int\limits_0^\infty\!{\rm d}u\,
   {\sin(\pi u)\over\pi u}\,{12\over(1-u)(3-u)(4-u)}\,
   \int\limits_{\lambda^2/m_\tau^2}^\infty\!{\rm d}\tau\,
   \widehat w_D(\tau)\,\bigg( {\tau m_\tau^2\over\Lambda_V^2}
   \bigg)^{-u} \,.
\end{equation}
As long as $\lambda>\Lambda_V$, the integral over $u$ is convergent
and can be performed first. Expressing the result in terms of an
effective coupling constant $G(a)$ as in (\ref{taurep2}), we find
that
\begin{equation}\label{Treg}
   T_{\rm reg}^{\rm circle}(m_\tau^2,\lambda^2) = 1
   + {1\over\beta_0}\,\int\limits_{\lambda^2/m_\tau^2}^\infty\!
   {\rm d}\tau\,\Omega_\tau(\tau)\,G\big(a(\tau m_\tau^2)\big) \,,
\end{equation}
where $\Omega_\tau(\tau)=\widehat w_D(\tau)$, and
\begin{equation}
   G(a) = \int\limits_0^\infty\!{\rm d}u\,{\sin(\pi u)\over\pi u}\,
   {12\over(1-u)(3-u)(4-u)}\,e^{-u/a} = G_3(a) \,;\quad
   \mbox{for}~a>0 \,,
\end{equation}
with $G_3(a)$ as defined in (\ref{G3fun}). Now the freedom in the
choice of the continuation to the region below the Landau pole
introduces an ambiguity of order $1/m_\tau^4$, since
$\Omega_\tau(\tau)\propto\tau$ for $\tau\ll 1$. An example of such a
continuation is provided by the expressions for $T_{\rm res}^{\rm
circle}(m_\tau^2)$ given in (\ref{taurep3}). If the IR cutoff is
kept, the ambiguity is reflected in the fact that
\begin{eqnarray}
   \lambda^2\,{{\rm d}\over{\rm d}\lambda^2}\,
   T_{\rm reg}^{\rm circle}(m_\tau^2,\lambda^2)
   &=& -{1\over\beta_0}\,G\big( a(\lambda^2) \big)\,
    {\lambda^2\over m_\tau^2}\,\Omega_\tau(\lambda^2/m_\tau^2)
    \nonumber\\
   &=& - {6 C_F\over\beta_0}\,G\big( a(\lambda^2) \big)\,\bigg(
    {\Lambda_V^2\over m_\tau^2} \bigg)^2 + O(1/m_\tau^6) \,,
\end{eqnarray}
which means that one has to add a nonperturbative contribution of
order $1/m_\tau^4$ to cancel the dependence on the cutoff.

It is again instructive to compare our result for $T_{\rm res}^{\rm
circle}(m_\tau^2)$ with the principle value of the Borel integral,
which according to Refs.~\cite{BBnew,xxx} can be written in the form
\begin{equation}
   T_{\rm Borel}(m_\tau^2) = 1 + {1\over\beta_0}\,
   \int\limits_0^\infty\!{\rm d}\tau\,W_\tau(\tau)\,
   F_1\big( a(\tau m_\tau^2) \big) + {1\over\beta_0}\,\mbox{Re}
   \int\limits_{-\tau_L}^0\!{\rm d}\tau\,W_\tau(\tau-i\epsilon) \,,
\end{equation}
with $\tau_L=\Lambda_V^2/m_\tau^2$. Using (\ref{T123}), we find that
\begin{eqnarray}\label{TBorcom}
   T_{\rm Borel}(m_\tau^2) - T_{\rm res}^{\rm circle}(m_\tau^2)
   &=& {1\over\beta_0}\,\mbox{Re} \int\limits_{-\tau_L}^{\tau_L}\!
    {\rm d}\tau\,W_\tau(\tau-i\epsilon) - {4\over\beta_0}\,
    \int\limits_{\tau_L}^\infty\!{\rm d}\tau\,\Omega_\tau(\tau)\,
    \bigg( {\tau_L\over\tau} - {\tau_L^3\over\tau^3} \bigg)
    \nonumber\\
   &=& {48 C_F\over\beta_0}\,\bigg( {\Lambda_V^2\over m_\tau^2}
    \bigg)^2 + O(1/m_\tau^6) \simeq 0.03 \,.
\end{eqnarray}
Although of order $1/m_\tau^4$, this difference is sizeable because
of the large numerical coefficient. For completeness, we also quote
the exact result for the renormalon ambiguity. It is
\begin{eqnarray}
   \Delta T_{\rm ren} &=& {4 C_F\over\beta_0}\,\bigg\{
    \tau_L + {47\over 6}\,\tau_L^2 - {\tau_L^3\over 3}
    - 4 L_2(\tau_L) - \bigg( 2\tau_L^3 - {\tau_L^4\over 3}
    \bigg)\,\ln\tau_L \nonumber\\
   &&\phantom{ {4 C_F\over\beta_0}\,\bigg\{ }
    + \bigg( 1 + {13\over 3}\,\tau_L - {5\over 3}\,\tau_L^2
    + {\tau_L^3\over 3} \bigg) (1-\tau_L)\,\ln(1-\tau_L) \bigg\} \,,
\end{eqnarray}
which for $\tau_L\ll 1$ reproduces (\ref{Tren}).

The discussion of this section illustrates that the choice of
implementing the IR regulator requires some amount of ingenuity. In
the first case, by introducing the regulator we have excluded a
contribution of order $1/m_\tau^2$ from the perturbative calculation
of $T(m_\tau^2)$, whereas in the second case we have found a better
regularization scheme, in which this contributions is of order
$1/m_\tau^4$. It is important that there is a physical argument which
favours the second method over the first one. The integration along
the circle in the complex plane involves larger momenta than the
integration along the cut, and thus less is eliminated by introducing
the IR cutoff. Clearly, the question arises whether there exists an
even better representation, in which the excluded contribution is of
order $1/m_\tau^6$ and thus of the same order as the renormalon
ambiguity. Although we cannot give a definite answer to this question
at present, we doubt that such a representation exists; we find it
suggestive that a linear integral representation of the form
(\ref{resum}), which would have this property, does not exist in the
physical region. As long as one does not succeed in constructing such
a regularized representation of the Borel integral, one has to take
seriously the possibility of nonperturbative corrections to $R_\tau$
of order $1/m_\tau^4$.

To summarize this section, we repeat that the new kind of
nonperturbative ambiguities encountered in our analysis is related to
the singular behaviour of the Borel integral for $u\to\infty$. We
have argued that this behaviour should be regulated by introducing an
IR cutoff. To the perturbative calculation one then has to add a
nonperturbative contribution in order to cancel the dependence on the
cutoff in the final result. For quantities defined in the euclidean
region, the power behaviour of the nonperturbative contribution is in
one-to-one correspondence with the size of the ambiguities due to IR
renormalons, which appear if the Borel integral is evaluated in
dimensional regularization without using an IR cutoff. This implies
that in these cases the singularities at infinity do not spoil the
applicability of the OPE. For the cases of correlation functions in
the physical region, however, we find that the singular behaviour at
infinity leads to ambiguities that may be larger than those indicated
by the positions of IR renormalon singularities. At least, we could
not find a cutoff regularization for which these ambiguities are of
the same order as those arising from renormalons. If this conjecture
is correct, it would imply that the OPE does not provide a consistent
framework for the analysis of nonperturbative corrections in the
physical region.

\section{Numerical analysis}
\label{sec:5}

We now turn to the numerical analysis of our results. In
Tables~\ref{tab:2} and \ref{tab:3} we show various approximations for
the quantities $R(m_\tau^2)$ and $T(m_\tau^2)$. Compare first the
results for $R(m_\tau^2)$ with those for $D(m_\tau^2)$ given in
Table~\ref{tab:1}. The corresponding perturbative series are the same
up to order $\alpha_s^2$. In the case of the euclidean correlation
function, the resummation of renormalon chains leads to an increase
of 3\% with respect to the two-loop result, whereas the value of the
physical correlation function is decreased by 1\%. This reflects the
smooth behaviour of the effective coupling constants $F_n(a)$ in the
IR region. As a consequence, after the resummation the result lies
significantly below the BLM prediction. Note that the difference
between the two resummations $R_{\rm res}^{(1)}$ and $R_{\rm
res}^{(2)}$ is numerically small, although larger than the renormalon
ambiguity $\Delta R_{\rm ren}$. We also quote the principle value of
the Borel integral, $R_{\rm Borel}$, which according to
(\ref{RBorcomp}) is very close to $R_{\rm res}^{(2)}$.

Consider now the results for $T(m_\tau^2)$, which has a larger
two-loop coefficient than $D(m_\tau^2)$ and $R(m_\tau^2)$. The
resummation of renormalon chains again leads to values that are lower
than the BLM approximation; however, as discussed in the preceding
sections there are significant differences between the various
resummations. Some of these differences are of order $1/m_\tau^2$ and
thus much larger than the renormalon ambiguity. We have also seen
that, for a physical reason, the resummation $T_{\rm res}^{\rm
circle}$ is to be preferred over the resummations $T_{\rm res}^{(1)}$
and $T_{\rm res}^{(2)}$. On the other hand, the difference between
$T_{\rm res}^{\rm circle}$ and the principle value of the Borel
integral constitutes a real ambiguity, which cannot be avoided on
physical grounds; at least at present we see no argument to prefer
one result over the other. This difference is an effect of order
$1/m_\tau^4$, which is still larger than the renormalon ambiguity.

\begin{table}[t]
\centerline{\parbox{15cm}{\caption{\label{tab:2}
Comparison of various approximations for the quantity $R(s)$ at
$s=m_\tau^2$.}}}
\vspace{0.5cm}
\centerline{\begin{tabular}{cccccccc}
\hline\hline
\rule[-0.2cm]{0cm}{0.7cm} $R_{\rm 1~loop}$ & $R_{\rm 2~loop}$ &
 $R_{\rm BLM}$ & $R_{\rm BLM^*}$ & $R_{\rm res}^{(1)}$ &
 $R_{\rm res}^{(2)}$ & $R_{\rm Borel}$ & $\Delta R_{\rm ren}$ \\
\hline
\rule{0cm}{0.6cm} 1.102 & 1.118 & 1.145 & 1.115 & 1.107 & 1.105 &
 1.105 & $-6.2\times 10^{-5}\phantom{-}$ \\
\hline\hline
\end{tabular}}
\vspace{0.5cm}
\end{table}

\begin{table}[t]
\centerline{\parbox{15cm}{\caption{\label{tab:3}
Comparison of various approximations for the quantity
$T(m_\tau^2)$.}}}
\vspace{0.5cm}
\centerline{\begin{tabular}{ccccccccc}
\hline\hline
\rule[-0.2cm]{0cm}{0.7cm} $T_{\rm 1~loop}$ & $T_{\rm 2~loop}$ &
 $T_{\rm BLM}$ & $T_{\rm BLM^*}$ & $T_{\rm res}^{(1)}$ &
 $T_{\rm res}^{(2)}$ & $T_{\rm res}^{\rm circle}$ &
 $T_{\rm Borel}$ & $\Delta T_{\rm ren}$ \\
\hline
\rule{0cm}{0.6cm} 1.102 & 1.155 & 1.213 & 1.238 & 1.151 & 1.097 &
 1.199 & 1.228 & $1.4\times 10^{-3}$ \\
\hline\hline
\end{tabular}}
\vspace{0.5cm}
\end{table}

\begin{figure}[htb]
   \vspace{0.5cm}
   \epsfysize=6cm
   \centerline{\epsffile{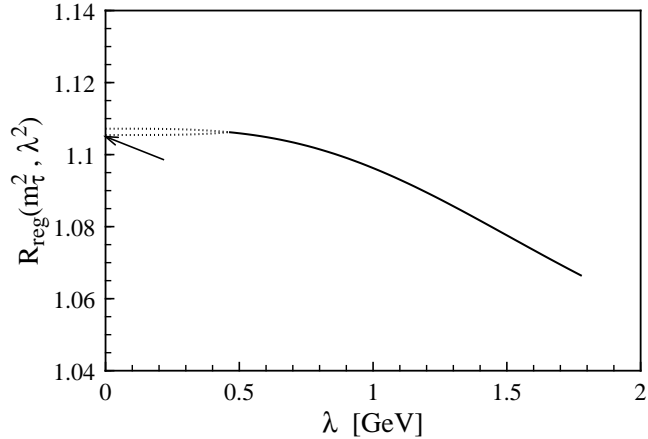}}
   \centerline{\parbox{13cm}{\caption{\label{fig:6}
The regularized integral $R_{\rm reg}(m_\tau^2,\lambda^2)$ as a
function of the cutoff scale. For values $\lambda<\Lambda_V\simeq
0.46$~GeV we define the integral by using the effective coupling
constants $F_1(a)$ (upper curve) and $F_2(a)$ (lower curve). The
arrow indicates the principle value of the Borel integral.}}}
\end{figure}

\begin{figure}[htb]
   \vspace{0.5cm}
   \epsfysize=6cm
   \centerline{\epsffile{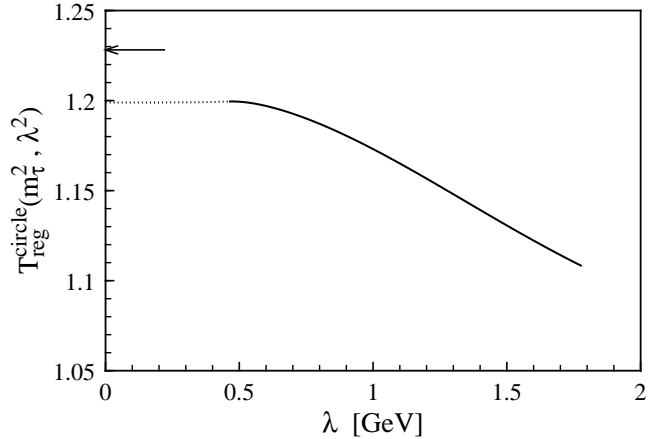}}
   \centerline{\parbox{13cm}{\caption{\label{fig:7}
The regularized integral $T_{\rm reg}^{\rm
circle}(m_\tau^2,\lambda^2)$ as a function of the cutoff scale. For
values $\lambda<\Lambda_V\simeq 0.46$~GeV we define the integral by
using the effective coupling constant $G_3(a)$. The arrow indicates
the principle value of the Borel integral.}}}
\end{figure}

Let us finally study the regularized integrals $R_{\rm
reg}(m_\tau^2,\lambda^2)$ and $T_{\rm reg}(m_\tau^2,\lambda^2)$ in
(\ref{Rreg}) and (\ref{Treg}) as a function of the cutoff scale. The
corresponding curves are shown in Figs.~\ref{fig:6} and \ref{fig:7}.
The regularized integrals are well defined down to a value
$\lambda=\Lambda_V\simeq 0.46$~GeV, where one reaches the Landau pole
in the running coupling constant. Below this value we define the
integrals by using one of the effective coupling constants
encountered in the preceding sections ($F_1(a)$ and $F_2(a)$ in the
case of $R(m_\tau^2)$, and $G_3(a)$ in the case of $T(m_\tau^2)$), so
that for $\lambda=0$ we recover the results given in
Tables~\ref{tab:2} and \ref{tab:3}. In each figure the arrow
indicates the principle value of the Borel integral. The variation of
the results in the low-momentum region, as well as the difference
between the resummations and the principle value of the Borel
integral, provide an estimate of the importance of nonperturbative
contributions.

\section{Conclusions}
\label{sec:6}

Generalizing the approach proposed in Ref.~\cite{part1}, we have
investigated the resummation of renormalon chain contributions in the
perturbative calculation of cross sections and inclusive decay rates,
and in particular for the ratios $R_{e^+ e^-}$ and $R_\tau$. Our
approach is equivalent to an all-order resummation of the terms
proportional to $\beta_0^{n-1}\alpha_s^n$ in the perturbative series
for these quantities. It provides a generalization of the BLM
scale-setting prescription \cite{BLM,LeMa}, in which terms of order
$\beta_0\,\alpha_s^2$ are absorbed by a redefinition of the scale in
the running coupling constant.

The discussion in Ref.~\cite{part1} was devoted to QCD Green
functions without external gluon fields and to euclidean correlation
functions of currents, which receive only virtual gluon corrections.
In the calculation of cross sections and inclusive decay rates, on
the other hand, both virtual and real gluons have to be considered.
As a consequence, in the expression for the resummed series, which
has the form of a one-dimensional integral over the running coupling
constant with some distribution function, there appears an effective
coupling constant, which is screened in the low-momentum region. Such
non-linear representations have also been derived, using a different
formalism, by Beneke and Braun \cite{BBnew}.

By summing an infinite set of diagrams our scheme reaches beyond
perturbation theory. In any finite-order perturbative calculation
nonperturbative effects are implicitly present due to low-momentum
contributions in Feynman diagrams, but are not visible as they are
exponentially small in the coupling constant. Yet perturbation theory
``knows'' about these contributions in the form of IR renormalons,
which make a perturbative series non Borel summable. This means that
attempts to resum the series will lead to unavoidable ambiguities. In
our scheme these ambiguities arise from the integration over the
Landau pole in the running coupling constant. They are a reminder of
the fact that perturbation theory is incomplete; any perturbative
calculation must be supplemented by nonperturbative contributions.
For the quantities considered in Ref.~\cite{part1}, the OPE provides
the framework for a consistent incorporation of nonperturbative
contributions. In this case the IR renormalon ambiguities can be
absorbed into the definition of other nonperturbative parameters,
such as the vacuum condensates. A new feature of the present analysis
is that it refers to quantities defined in the physical region (i.e.\
for time-like momenta), for which the applicability of the OPE is at
least questionable. We have found a new source of nonperturbative
ambiguities in the definition of the resummed series, which are not
related to IR renormalons. These ambiguities are parametrically
larger than the IR renormalon ambiguities and thus cannot be dealt
with in the standard way.

The appearance of this new source of ambiguity is surprising and is
the most striking observation of our analysis. In order to understand
the origin and size of this effect, we have regularized the singular
behaviour of the Borel integral by introducing an IR cutoff. We find
that in the euclidean region there is a one-to-one correspondence
between the power-like dependence on this cutoff and the position of
IR renormalons, which establishes the link between the OPE and the
structure of the singularities in the Borel plane. However, in the
physical region we did not succeed to construct a cutoff
regularization with this property. In other words, for the
regularization schemes that we have investigated the cutoff
dependence does not match with the position of IR renormalons.
Although we cannot prove at present that no better regularization
exists, we cannot exclude the possibility that the ambiguities
encountered in our analysis of $R_{e^+ e^-}$ and $R_\tau$ hinder the
standard application of the OPE in the physical region. This would
imply that to the resummed perturbative series one would have to add
nonperturbative power corrections that are not related to vacuum
expectation values of local, gauge-invariant operators. The values of
these nonperturbative terms would depend on the resummation
prescription chosen for the perturbative series. It is not obvious to
us which theoretical framework could be employed to accommodate these
terms.

It is of great importance to settle the question if this pessimistic
possibility can be ruled out. Let us illustrate some phenomenological
implications with two examples. We have mentioned in the introduction
that the inclusive $B\to X\,e\, \bar\nu_e$ decay rate is subject to
large two-loop corrections of order $\beta_0\,\alpha_s^2$, and that
the corresponding BLM scale is very low. In the case of $b\to u$
transitions, for instance, one finds $\mu_{\rm BLM}\simeq 0.07\,m_b$
in the $\overline{\rm MS}$ scheme \cite{LSW}. This observation casts
doubt on the reliability of the perturbative expansion and calls for
an analysis of higher-order terms in the series. It is thus very
interesting to investigate the resummation of renormalon chains in
this case. The principle value of the Borel integral corresponding to
the perturbative series for the inclusive semileptonic decay rate has
recently been calculated by Ball et al.\ \cite{Patri}. However, based
on our analysis we expect that also in this case the result of the
resummation will not be unique, i.e.\ there will be ambiguities of
order $1/m_b^2$ or even $1/m_b$. In the latter case this would limit
the accuracy with which one can extract the elements $V_{ub}$ and
$V_{cb}$ of the Cabibbo-Kobayashi-Maskawa matrix from the analysis of
inclusive decays of $B$ mesons.

Our second example concerns the quantity $R_\tau$, measurements of
which have been suggested as a way to extract with high precision the
strong coupling constant $\alpha_s(m_\tau^2)$ \cite{Rtau1,Rt5},
\cite{Rt1}--\cite{Rt4}. From the most recent analysis of the
experimental data, one has obtained the value $R_\tau=3.638\pm 0.017$
and extracted $\alpha_s(m_\tau^2)=0.367\pm 0.018$ \cite{Duflot}. The
theoretical uncertainty in this result is quoted as
$\delta\alpha_s(m_\tau^2)=0.014$ \cite{Rt5}. To achieve such a high
precision, it is essential to rely on the standard form of the OPE,
in which the magnitude of nonperturbative contributions is linked to
the structure of renormalon singularities in the Borel plane. Using
the SVZ approach \cite{SVZ} to relate these nonperturbative
contributions to vacuum condensates, one finds that the contribution
of the gluon condensate, which is proportional to $\langle\alpha_s
G^2\rangle/m_\tau^4$, is suppressed by two additional powers of
$\alpha_s$ (it is absent in the large-$\beta_0$ limit considered
here), so that the dominant nonperturbative effects appear at order
$1/m_\tau^6$ \cite{Rtau1}. Several authors have questioned this
assumption and investigated, with different conclusions, whether
$R_\tau$ could receive larger nonperturbative corrections. Both IR
and UV renormalons have received main attention in these studies
\cite{reno5}, \cite{West}--\cite{Domi}. In the present work we have
identified a new kind of nonperturbative effects, which are not
related to vacuum condensates. We have seen that there exist
different ways to resum renormalon chain contributions. Some of these
resummations are disfavoured for physical reasons, but others cannot
be distinguished on physical grounds. In the case of $R_\tau$, we see
no physical argument to prefer one of the two resummations $T_{\rm
res}^{\rm circle}$ and $T_{\rm Borel}$, which differ by terms of
order $1/m_\tau^4$. More generally, if the Borel integral is
regulated by introducing an IR cutoff, one is forced to add a
nonperturbative contribution of order $1/m_\tau^4$ to cancel the
cutoff dependence of the perturbative calculation.

To illustrate the phenomenological impact of our results, let us
compare the various predictions for the perturbative contribution to
$R_\tau$. We write
\begin{equation}
   R_\tau = 3\,(|\,V_{ud}|^2 + |\,V_{us}|^2)\,S_{\rm EW}\,
   \Big\{ 1 + \delta_{\rm EW} + \delta_{\rm pert}
   + \delta_{\rm nonpert} \Big\} \,,
\end{equation}
where $S_{\rm EW}\simeq 1.0194$ and $\delta_{\rm EW}\simeq 0.0010$
are known electroweak corrections \cite{Sirl,BraLi}, $\delta_{\rm
pert}$ is the perturbative correction, and $\delta_{\rm nonpert}$
denotes the sum of all power-suppressed corrections including
quark-mass effects, which have been estimated to be $\delta_{\rm
nonpert}=-0.019\pm 0.007$ \cite{Rtau1}. Accepting this number, one
obtains $\delta_{\rm pert}^{\rm exp}=0.21\pm 0.01$ from the
experimental measurements. It is conventional to consider the
perturbative correction as a function of the coupling constant
$\alpha_s(m_\tau^2)$ defined in the $\overline{\rm MS}$ scheme. We
use the known exact values of the coefficients up to order
$\alpha_s^3$ \cite{Tals3a,Tals3b} to correct our result for
contributions which are subleading in the large-$\beta_0$ limit and
thus not included in our approach. This leads to
\begin{equation}\label{dpert}
   \delta_{\rm pert}[\alpha_s(m_\tau^2)]
   = T[\alpha_s(m_\tau^2)] - 1 + 0.0831\,{\alpha_s(m_\tau^2)\over\pi}
   - 2.4115\,\bigg( {\alpha_s(m_\tau^2)\over\pi} \bigg)^2 \,,
\end{equation}
where for $T$ we may substitute one of the resummations $T_{\rm
res}^{\rm circle}$ or $T_{\rm Borel}$, which depend on
$\alpha_s(m_\tau^2)$ through the QCD scale parameter $\Lambda_V$.

\begin{figure}[htb]
   \vspace{0.5cm}
   \epsfysize=6cm
   \centerline{\epsffile{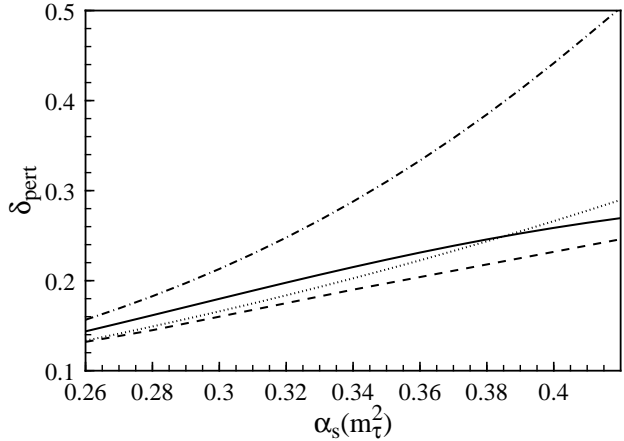}}
   \centerline{\parbox{13cm}{\caption{\label{fig:8}
Theoretical predictions for the quantity $\delta_{\rm pert}$: this
work, eq.~(\protect\ref{taurep3}) (solid line); principle value of
the Borel integral, from \protect\cite{xxx} (dash-dotted line);
resummation of Le~Diberder and Pich, from \protect\cite{Rt5} (dashed
line); exact order-$\alpha_s^3$ result (dotted line).}}}
\end{figure}

In Fig.~\ref{fig:8}, we compare our results with the resummation of
Pich and Le~Diberder \cite{Rt5}, as well as with the exact
order-$\alpha_s^3$ expression. There are significant differences
between the various predictions. For instance, with the central
``experimental'' value $\delta_{\rm pert}=0.21$ one would obtain
$\alpha_s(m_\tau^2)\simeq 0.30$ using the principle value
resummation, $\alpha_s(m_\tau^2)\simeq 0.34$ using our resummation
$T_{\rm res}^{\rm circle}$, and $\alpha_s(m_\tau^2)\simeq 0.37$ using
the resummation of Le~Diberder and Pich. If $\delta_{\rm pert}$ is
larger, the differences even increase. We conclude that the
theoretical uncertainty in the perturbative contribution to $R_\tau$
is probably larger than estimated in \cite{Rt5}. A reasonable value
seems to be
\begin{equation}
   \delta\alpha_s(m_\tau^2) \simeq 0.05 \,.
\end{equation}

\subsection*{Acknowledgements}

While this paper was in writing I was informed about the work
\cite{xxx}, which partially overlaps with the present one. I
gratefully acknowledge exchange of manuscripts and many useful
discussions with P.~Ball and M.~Beneke. For further clarifying
discussions I like to thank G.~Altarelli, B.~Gavela, M.~Jamin,
P.~Nason, O.~P\`ene and A.~Pich.

\newpage
\section*{Appendix}

Beneke and Braun have developed a different way to resum renormalon
chains \cite{BBnew} (see also \cite{xxx}). They start from the
integral representation (\ref{BBrepr}) for the Borel transform and
show that the Borel integral performed above the real $u$-axis is
equivalent to
\begin{equation}\label{nonlin}
   S_{\rm Borel}(M^2) = 1 + {1\over\beta_0}\,
   \int\limits_0^\infty\!{\rm d}\tau\,W(\tau)\,
   F_1\big(a(\tau M^2)\big) + {1\over\beta_0}\,
   \int\limits_{-\tau_L}^0\!{\rm d}\tau\,W(\tau-i\epsilon) \,,
\end{equation}
where $F_1(a)$ has been defined in (\ref{Fdef}). In this approach,
the renormalon ambiguity is given by
\begin{equation}\label{Srendef2}
   \Delta S_{\rm ren} = {1\over\pi\beta_0}\,\mbox{Im}
   \int\limits_{-\tau_L}^0\!{\rm d}\tau\,W(\tau-i\epsilon) \,.
\end{equation}

The function $W(\tau)$ can be calculated by performing a one-loop
calculation with a finite gluon mass $m_g^2=\tau M^2$. Its relation
to the distribution function $\widehat w(\tau)$ of our approach has
been given in (\ref{Wtwtau}). Using this relation, one can recover
our linear representation (\ref{resum}) from the non-linear
representation (\ref{nonlin}) of Beneke and Braun. Inserting
(\ref{Wtwtau}) into (\ref{nonlin}) and changing the order of
integration (which is allowed since the integrals converge), one
finds that
\begin{equation}
   S_{\rm Borel}(M^2) = 1 + {1\over\beta_0}\,
   \int\limits_0^\infty\!{\rm d}\tau\,\widehat w(\tau)\,
   H\big(a(\tau M^2)\big) \,,
\end{equation}
where
\begin{equation}
   H(a) = \int\limits_0^\infty\!{{\rm d}z\over(1+z)^2}\,
   {1\over\pi}\mbox{arccot}\,\bigg( {\ln z+1/a\over\pi} \bigg)
   + {1\over 1-e^{-1/a}-i\epsilon} - 1 \,.
\end{equation}
After an integration by parts in the first term this can be written
in the form
\begin{equation}
   H(a) = {1\over 1-e^{-1/a}-i\epsilon}
   - \int\limits_{-\infty}^{+\infty}\!{\rm d}x\,
   {1\over(x^2+\pi^2)(1+e^{x-1/a})} \,.
\end{equation}
The $x$-integral can be performed using Chauchy's theorem. There are
simple poles at $x=\pm i\pi$ and $x=1/a+(2 n+1)\,i\pi$. We find
\begin{eqnarray}
   \int\limits_{-\infty}^{+\infty}\!{\rm d}x\,
   {1\over(x^2+\pi^2)(1+e^{x-1/a})} &=& {1\over 1-e^{-1/a}}
    + \sum_{n=0}^\infty\,\bigg( {1\over 1/a + 2n\,i\pi}
    - {1\over 1/a + 2(n+1)\,i\pi} \bigg) \nonumber\\
   &=& {1\over 1-e^{-1/a}} + a \,,
\end{eqnarray}
which leads to
\begin{equation}
   H(a) = a + i\pi\,\delta(1/a) = {1\over 1/a - i\epsilon}
   = {1\over\ln\tau - \ln\tau_L - i\epsilon} \,.
\end{equation}
This shows the equivalence of the two methods in the euclidean
region.

Using the relation (\ref{Wtwtau}), let us derive some properties of
the function $W(t)$. We note that $W(t)$ is analytic in the complex
$t$-plane with a branch cut along the negative axis. For the
discontinuity across the cut, we obtain
\begin{equation}
   {1\over 2\pi i}\,\Big[ W(\tau+i\epsilon) - W(\tau-i\epsilon)
   \Big] = -\Theta(-\tau)\,{{\rm d}\over{\rm d}\tau}\,\Big[
   \tau\,\widehat w(-\tau) \Big] \,.
\end{equation}
Using this result, we find that our definition of the renormalon
ambiguity in (\ref{DeltaS}) is in fact equivalent to
(\ref{Srendef2}), since
\begin{equation}
   \Delta S_{\rm ren} = {1\over\pi\beta_0}\,\mbox{Im}\,
   \int\limits_{-\tau_L}^0\!{\rm d}\tau\,W(\tau-i\epsilon)
   = {\tau_L\over\beta_0}\,\widehat w(\tau_L) \,.
\end{equation}

For the cases of $R_{e^+ e^-}(s)$ and $R_\tau$, the functions $W_{e^+
e^-}(\tau)$ and $W_\tau(\tau)$ have been given in (\ref{WeewD}) and
(\ref{wtaufun}). These results agree with the corresponding
expressions derived in Ref.~\cite{xxx}. For completeness, let us also
quote the result for the function $W_D(t)$ in the case of the
$D$-function. Starting from relation (\ref{Wtwtau}) and using the
result for the distribution function $\widehat w_D(\tau)$ given in
(\ref{wDfun}), we obtain
\begin{eqnarray}
   W_D(\tau) &=& 8 C_F\,\Bigg\{ {7\over 2} - {5\tau\over 2}\,
    \ln{1+\tau\over\tau} + \bigg( {1\over 2\tau} - 2 \bigg)\,
    \ln(1+\tau) + 2\,L_2(-\tau) + \tau\,L_2(-1/\tau) \nonumber\\
   &&\mbox{}- (1-2\tau)\,\int\limits_0^1\!{\rm d}x\,
    {\ln x\over 1+x}\,\ln{x(x+\tau)\over(1+x\tau)} \Bigg\}
    \nonumber\\
   &=& 8 C_F\,\bigg\{ 4 - 3\zeta(3) + \bigg( 6\zeta(3) - {17\over 4}
    + {3\over 2}\,\ln\tau \bigg)\,\tau + O(\tau^2) \bigg\} \,.
\end{eqnarray}
The first two terms in the expansion of this function have been
calculated in Ref.~\cite{xxx}; the exact expression is new.

\end{document}